\documentclass[universe,article,accept,pdftex,moreauthors]{Definitions/mdpi}
\firstpage{1}
\pubvolume{12}
\issuenum{7}
\articlenumber{214}
\pubyear{2026}
\copyrightyear{2026}
\externaleditor{Stefano Vercellone} 
\datereceived{25 May 2026}
\daterevised{23 June 2026} 
\dateaccepted{25 June 2026}
\datepublished{16 July 2026 }

\usepackage{longtable}

\usepackage{xcolor}
\newcommand{\rev}[1]{\textcolor{black}{#1}}

\Title{Testing [O~{\sc \textbf{ii}}] $\lambda3727$ as a Star Formation Rate Tracer in Quasar Host Galaxies}

\Author{{Xiaotong Feng}~$^{1,2,}$*\href{https://orcid.org/0000-0003-0174-5920}{\orcidicon}, {Xue-Bing Wu} $^{1,2}$\href{https://orcid.org/0000-0002-7350-6913}{\orcidicon}, Yuming Fu $^{3,4}$\href{https://orcid.org/0000-0002-0759-0504}{\orcidicon}, Yuxuan Pang $^{5}$\href{https://orcid.org/0009-0005-3823-9302}{\orcidicon}, Rui Zhu $^{1,2}$\href{https://orcid.org/0000-0002-0792-2353}{\orcidicon} and Huimei Wang $^{1,2}$\href{https://orcid.org/0000-0001-8803-0738}{\orcidicon}}

\AuthorNames{Xiaotong Feng, Xue-Bing Wu, Yuming Fu, Yuxuan Pang, Rui Zhu and Huimei Wang}

\address{%
$^{1}$ \quad Department of Astronomy, School of Physics, Peking University, Beijing 100871, China
\\
$^{2}$ \quad Kavli Institute for Astronomy and Astrophysics, Peking University, Beijing 100871, China\\
$^{3}$ \quad Leiden Observatory, Leiden University, Einsteinweg 55, 2333 CC Leiden, The Netherlands
\\
$^{4}$ \quad Kapteyn Astronomical Institute, University of Groningen, P.O. Box 800, 9700 AV Groningen, The Netherlands\\
$^{5}$ \quad School of Astronomy and Space Science, University of Chinese Academy of Sciences (UCAS), Beijing~100049,~China
}

\corres{Correspondence: xtfeng@pku.edu.cn}

\abstract{
The [O~{\sc ii}] $\lambda3727$ emission line is a widely used star formation rate (SFR) tracer.
However, its application to type I quasars is not straightforward, because the line can be affected by dust extinction, metallicity and contamination from the AGN narrow-line region (NLR).
We test the reliability of [O~{\sc ii}] SFRs using a sample of 202 SDSS and PG quasars, by comparing [O~{\sc ii}] SFRs and reference far-infrared (FIR) SFRs derived from multiwavelength SED decomposition.
We measure [O~{\sc ii}], [O {\sc iii}], and narrow Balmer emission lines by spectral fitting. Then, we calculate [O~{\sc ii}] SFRs after correcting dust extinction and metallicity.
We then compare these SFRs with the FIR SFRs, both with and without subtracting the AGN contribution estimated from [O {\sc iii}].
After this correction, the median offset between \linebreak {}[O~{\sc ii}] and FIR SFRs is $-0.20\pm0.72$ dex for the full analysis sample and $-0.17\pm0.69$ dex for sources with S/N $>5$ in both [O~{\sc ii}] and [O {\sc iii}]. Without subtracting the AGN contribution, the corresponding offsets are $0.00\pm0.69$ and $0.12\pm0.66$ dex. We conclude that [O~{\sc ii}] is useful as a statistical SFR tracer for quasar host galaxies, but individual objects still require careful treatment of AGN contamination, extinction, metallicity, aperture effects, and redshift-dependent systematics.
}


\begin{document}

\section{Introduction}\label{sec:introduction}

Understanding the connection between supermassive black hole growth and star formation of the quasar host galaxies requires reliable measurements of the host-galaxy star formation rate (SFR).
Far-infrared (FIR) emission is a reliable tracer of the star formation of quasar host galaxy, but deep FIR observations are not available for many large quasar samples.
Optical spectra, in contrast, are available for large numbers of quasars from surveys such as the Sloan Digital Sky Survey (SDSS).
If a robust optical emission-line SFR tracer can be applied to quasars, it would provide an efficient and complementary way to study star formation in quasar host galaxies.

\rev{{FIR} 
and optical emission-line indicators provide complementary measurements of star formation. When well-sampled FIR photometry and a reliable SED decomposition are available, FIR emission from the cold-dust component provides a useful reference SFR because it traces star formation that is obscured by dust. However, deep FIR observations are not available for most large quasar samples, and the relatively low angular resolution of FIR data can introduce blending and source-confusion uncertainties. In addition, separating the dust heated by star formation from the dust heated by AGN requires multiwavelength~modeling.}

\rev{In contrast, optical spectra are available for large quasar samples from surveys such as SDSS.}
The [O~{\sc ii}] $\lambda3727$ emission line is one of the most widely used optical SFR tracers for star-forming galaxies.
\rev{The [O~{\sc ii}] $\lambda3727$ line is produced in H {\sc ii} regions ionized by young massive stars and can be measured directly from an optical spectrum over a broad redshift range \citep{1998ARA&A..36..189K,2004AJ....127.2002K}.}
Although the H$\alpha$ emission line is a more straight tracer, its wavelength is easy to be redshifted out of the observed optical wavelength range or not covered by the available spectra.
[O~{\sc ii}] $\lambda3727$ has a shorter wavelength, \rev{making it accessible at higher redshift than H$\alpha$ in optical spectroscopic surveys. Therefore, [O~{\sc ii}] is not expected to be more reliable than a carefully derived FIR SFR for individual quasars; its main advantage is that it can provide an observationally efficient SFR estimate for large spectroscopic samples when deep FIR data are unavailable. The purpose of this work is not to replace FIR SFRs, but to test whether [O~{\sc ii}] can serve as a useful statistical SFR tracer for quasar host galaxies and to identify its main limitations.}
Unlike H$\alpha$, [O~{\sc ii}] is sensitive to dust attenuation, gas-phase metallicity, and ionization state.
These dependencies introduce substantial scatter, so we need careful calibration to use [O~{\sc ii}] as a quantitative SFR tracer \citep{2004AJ....127.2002K,2006ApJ...642..775M}.

But applying [O~{\sc ii}] to AGNs and quasars is more challenging.
In addition to star-forming H II regions, [O~{\sc ii}] \rev{can also arise} from AGN narrow-line region (NLR).
Therefore, the total [O~{\sc ii}] luminosity of a quasar is not a pure tracer of star formation.
\rev{The} [O~{\sc ii}] SFR may be overestimated if we do not subtract the AGN contribution.
This difficulty is especially important for type-I quasars.
Because type-I quasars have a strong AGN continuum, broad emission lines, Fe emission, and narrow-line emission, they can influence the measurement of the emission features from the host galaxies.

\citet{2019ApJ...882...89Z} recalibrated the relation between [O~{\sc ii}] luminosity and SFR using a large sample of star-forming galaxies.
They used a photoionization model to estimate how much the [O~{\sc ii}] emission comes from the AGN NLR region.
By subtracting the contribution from AGN NLR, the [O~{\sc ii}] emission from host galaxies can be separated.
After that, they found the corrected [O~{\sc ii}] SFRs in good agreement with D4000 SFRs for a large sample of SDSS narrow-line AGNs.
The median difference between these two SFRs is small, about 0.055 dex, though the scatter is large, about 0.335 dex.
However, it is still unclear whether this method works well for luminous type-I quasars.
In these quasars, the [O~{\sc ii}] emission can be strongly affected \rev{by the} luminous central AGN.

In this work, we investigate the reliability of [O~{\sc ii}] $\lambda3727$ as an SFR tracer for quasar host galaxies.
We use a sample of 202 quasars, including SDSS quasars and PG quasars, with both optical spectra and FIR SFRs.
The FIR SFRs are derived from multiwavelength SED decomposition and provide an independent estimation of SFR.
We refit the SDSS spectra and newly obtained optical spectra for PG quasars to measure [O~{\sc ii}] $\lambda3727$, [O~{\sc iii}] $\lambda5007$ and narrow Balmer lines.
We then derive [O~{\sc ii}] SFRs after correcting \rev{by} dust attenuation, metallicity and the AGN NLR contribution.
By comparing the [O~{\sc ii}] SFRs with FIR SFRs, we find that [O~{\sc ii}] is useful as a statistical SFR tracer of quasar host galaxies, but \rev{it} needs careful treatment of AGN contamination for individual objects.
We also examine how the SFR offset depends on AGN activity, the infrared AGN fraction, and redshift, in order to identify possible sources of systematic uncertainty.
The goal of this paper is not to investigate the FIR SFRs or dust properties themselves. Instead, the FIR SFRs are adopted as reference measurements for testing the [O~{\sc ii}] emission line method.
The derivation, model dependence, and astrophysical implications of the FIR SFRs are discussed in a separate companion study.

This paper is organized as follows. Section~\ref{sec:sample} describes the quasar sample and the FIR reference SFRs. Section~\ref{sec:spectroscopy} presents the optical spectroscopic observations, spectral fitting, and emission-line measurements. Section~\ref{sec:oii_sfr} describes the derivation of [O~{\sc ii}] SFRs. Section~\ref{sec:compareo2firsfr} compares the [O~{\sc ii}] SFRs with FIR SFRs and discusses the reliability and limitations of the [O~{\sc ii}] method. We summarize our main results in Section~\ref{sec:summary}.

\section{Sample}
\label{sec:sample}

\subsection{Sample Selection}
\label{subsec:sampleselection}

Our quasar sample consists of two subsamples: one is selected from the SDSS DR7 quasar catalog \citep{2009ApJS..182..543A,2010AJ....139.2360S, 2000AJ....120.1579Y} and the other is from the PG quasar sample \citep{BG1992, SG1983}.
Since the main goal of this work is to examine the reliability of the [O~{\sc ii}] $\lambda3727$ emission line as a SFR tracer of quasar host galaxies and compare it with independent FIR SFRs derived from multiwavelength SED decomposition, we require a quasar sample that has two parts of data: optical spectra covering the [O~{\sc ii}] $\lambda3727$ and [O {\sc iii}] $\lambda5007$ emission lines, enough FIR photometric to constrain the FIR SED fitting to estimate FIR SFR.

For the SDSS DR7 quasar catalog, we restrict the redshift range to $0.02<z<0.8$ and then cross-match it with public {Herschel} 
photometric catalogs from the NASA/IPAC Infrared Science Archive {(IRSA)}
\endnote{\url{https://irsa.ipac.caltech.edu/}~{(accessed on 24 June 2026).}
}.
We require each source to be detected in at least two of the six {Herschel} 
bands (PACS 70, 100, and 160~{$\upmu$}
m; SPIRE 250, 350, and 500~{$\upmu$}m), so that the FIR SED can be meaningfully constrained.
This selection results in 120 SDSS quasars with both optical spectra and FIR constraints.

PG quasars have been well studied and have plenty of spectroscopic and multiwavelength photometric data. In particular, the infrared photometry of 87 PG quasars has been compiled and modeled by \citet{shangguan2018}. These are optically bright, low-redshift ($z<0.5$) quasars. Five of the 87 PG quasars overlap with the 120 SDSS quasars selected above. After removing these duplicates, the final sample contains 202 unique~quasars.

\subsection{Reference FIR Star Formation Rates}

In this work, the FIR SFRs are used as reference values for [O~{\sc ii}] SFRs to compare. A detailed analysis of the submillimeter observations, multiwavelength SED fitting, cold-dust templates, and host galaxy properties will be presented in a companion paper. Here, we only summarize the aspects that are required for the present comparison. The reference FIR SFRs were derived from multiwavelength SED decomposition of stellar, dust, AGN, and radio components. The FIR luminosity used for the SFR estimation is calculated from the cold-dust component integrated over $8$--$1000~\upmu{\rm m}$ in order to reduce contamination from AGN-heated dust.
The adopted conversion is
\begin{equation}
{\rm SFR}_{\rm FIR}~(M_\odot~{\rm yr}^{-1}) =
3.02 \times 10^{-44} L_{\rm FIR}^{\rm cold}~({\rm erg~s}^{-1}),
\end{equation}
where $L_{\rm FIR}^{\rm cold}$ is the infrared luminosity of the cold-dust
integrated over $8$--$1000~\upmu{\rm m}$.
This conversion corresponds to the FIR SFR calibration of \citet{1998ARA&A..36..189K}, rescaled to a Kroupa~\citep{2001MNRAS.322..231K} initial mass function (IMF).
Throughout this paper, all SFRs and stellar masses are applied to the same Kroupa IMF scale.
The present paper does not attempt to interpret the FIR SFRs themselves, but uses them only to evaluate the reliability and limitations of [O~{\sc ii}] SFR~tracer.

\section{Spectroscopic Observations and Measurements}
\label{sec:spectroscopy}

\subsection{[O~{\sc ii}] $\lambda3727$ Measurements for the SDSS Quasar Sample}
\label{subsec:sdsso2}

For the 120 SDSS quasars in our sample, the catalog of \citet{shen11} provides measurements of the continuum and a number of broad and narrow emission lines. These include the monochromatic luminosities at 1350, 3000, and 5100~\AA, the broad and narrow components of H$\alpha$ and H$\beta$, and the narrow-line luminosities of [N {\sc ii}] $\lambda\lambda6548,6584$, [S {\sc ii}] $\lambda\lambda6717,6731$, and [O {\sc iii}] $\lambda\lambda4959,5007$. However, the [O~{\sc ii}] $\lambda3727$ line required for this work is not included in the catalog. We therefore refit the SDSS optical spectra to measure the [O~{\sc ii}] $\lambda3727$ luminosity.

The spectral fitting was performed with QSOFITMORE \citep{yuming_fu_2021_5810042}, a quasar spectral-fitting package designed to model both continuum and emission-line components.
First, we correct the spectra for Galactic extinction using the dust map of \citet{1998ApJ...500..525S}.
Second, since the redshifts of the quasars are known, all the spectra are shifted to the rest frame.
Third, we decompose each spectrum into host galaxy and quasar components using principal component analysis (PCA).
Fourth, we fit the quasar continuum using Fe emission templates \citep{BG1992,2001ApJS..134....1V,2007ApJ...662..131S}, a Balmer continuum component, and polynomial terms, using continuum windows that are relatively free from strong emission-line contamination.
After subtracting the best-fit continuum, we fit the emission-line components. The H$\beta$ and H$\alpha$ regions are fitted simultaneously with their neighboring narrow lines, including [O {\sc iii}] $\lambda\lambda4959,5007$, [N {\sc ii}] $\lambda\lambda6548,6584$, and [S {\sc ii}] $\lambda\lambda6717,6731$.
Both H$\beta$ and H$\alpha$ are modeled with broad and narrow components, while the forbidden lines are modeled as narrow components.
For [O {\sc iii}] $\lambda\lambda4959,5007$, an additional blue-wing component is included when required.
Each component is modeled with Gaussian profiles, the FWHM of the narrow components is constrained to be less than $800~{\rm km~s^{-1}}$.
The uncertainty of each fitted quantity is estimated from 50 Monte Carlo realizations.

\begin{figure}[H]
\centering
\includegraphics[width=\textwidth]{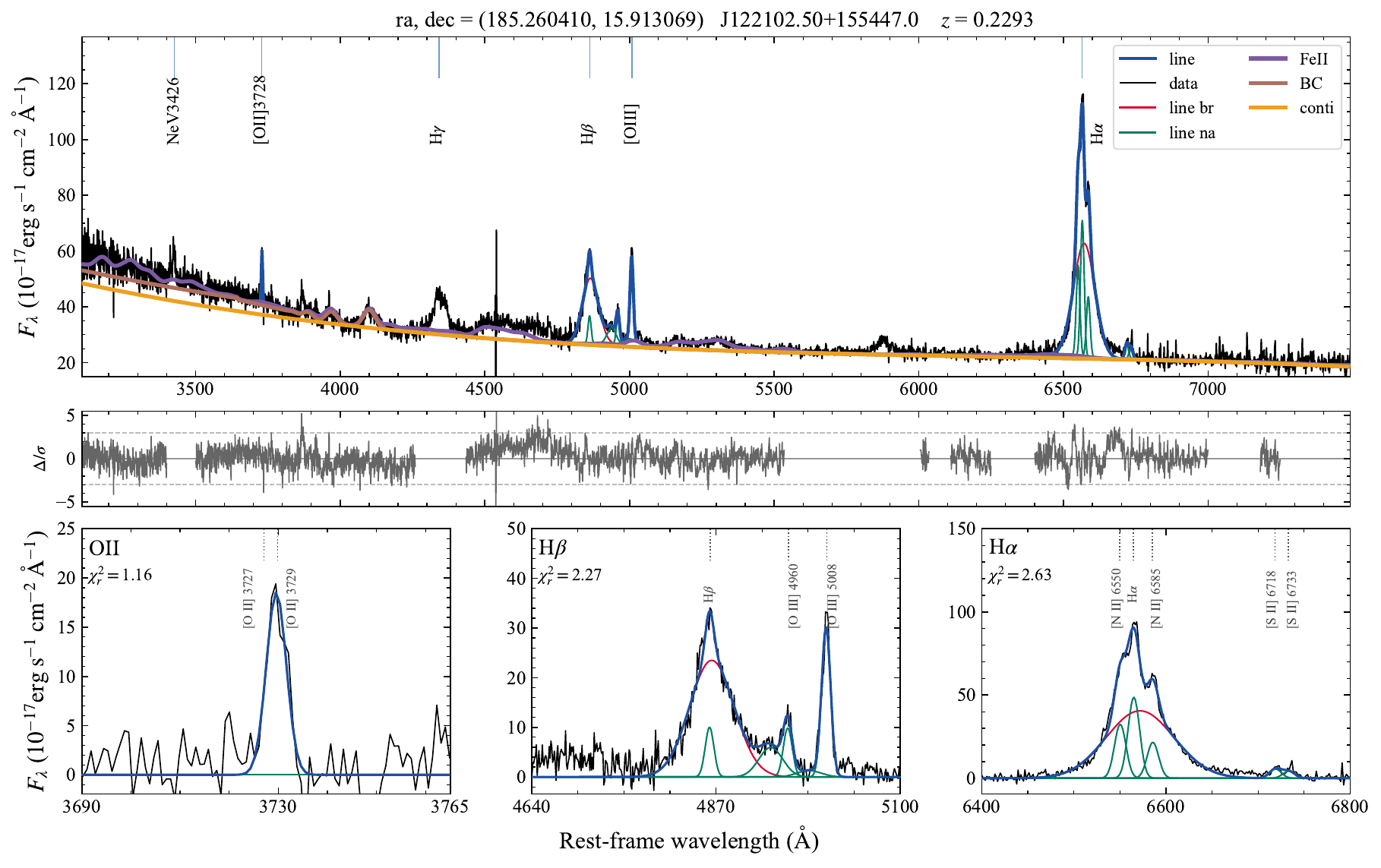}
\caption{\rev{{Example} 
of the spectral fitting for the SDSS quasar J122102.50+155447.04 using QSOFITMORE. The black curve shows the observed spectrum.
{The blue curve shows the total fitted emission-line model. The purple, brown and orange curves represent the Fe~{\sc ii} emission, the Balmer continuum, and the underlying continuum, respectively.} The red and green curves denote the broad and narrow emission-line components, respectively. The unresolved [O~{\sc ii}] feature is modeled with a single Gaussian profile and the broad H$\beta$ profile is represented by one broad Gaussian component. For [O {\sc iii}] $\lambda\lambda4959,5007$, the green wide components on the blue side of the line profiles indicate the [O {\sc iii}] outflow-wing components. The middle panel shows the residuals normalized by the spectral uncertainty ($(F_{\rm data}-F_{\rm model})/\sigma$) of the continuum and emission lines we fitted. It does not show the residuals of the emission lines we did not fit.}}
\label{fig:J122102spectra}
\end{figure}

Figure \ref{fig:J122102spectra} shows an example of the spectral fitting for one SDSS quasar in our sample, J122102.50{+}
{155447.04.} 
In the figure, the red curves represent the broad emission-line components, while the green curves represent the narrow emission-line components.
\rev{For [O {\sc iii}] $\lambda\lambda4959,5007$, the green wide components on the blue side of the line profiles indicate the [O {\sc iii}] outflow-wing components. Although the main analysis adopts a single-Gaussian approximation for the unresolved [O~{\sc ii}] doublet and one broad H$\beta$ component, we assess the impact of these choices using an alternative fitting scheme for a high-quality validation subsample in Section~\ref{subsec:comparefittingmethod}. Residuals normalized by the spectral uncertainty are also shown to illustrate the fit quality.}

The narrow emission line luminosity results of the SDSS quasars are listed in \linebreak {}Appendix~{\ref{app:measurements}}~
Table~\ref{tab:115sdssspecresult}.
The table includes [O~{\sc ii}], [O {\sc iii}], H$\alpha$, and H$\beta$ measurements for 115 SDSS quasars. The remaining five SDSS quasars overlap with the PG quasar sample and are listed together with the PG quasars in Section~\ref{subsec:pgo2}.
Because the SDSS DR7 spectra cover only 3800--9200~\AA, the H$\alpha$ emission line is shifted out for sources at $z\gtrsim0.4$.
So the H$\alpha$ luminosities are not available for these objects.
For several quasars, the H$\alpha$ or H$\beta$ emission is dominated by the broad component, so the corresponding narrow component is difficult to fit.
In addition, some spectra have low signal-to-noise ratios or have low spectral quality in specific wavelength regions, so we do not get reliable measurements of certain emission lines.
These unreliable measurements will be excluded from the subsequent analysis according to their signal-to-noise ratios.

\subsection{[O~{\sc ii}] $\lambda3727$ Measurements for the PG Quasar Sample and Observations}
\label{subsec:pgo2}

Among the 87 PG quasars, 31 have available SDSS spectra.
We fit these spectra with QSOFITMORE using the same procedure described in Section~\ref{subsec:sdsso2}.
The resulting narrow-line luminosities of [O~{\sc ii}], [O {\sc iii}], H$\alpha$, and H$\beta$ are listed in Appendix \ref{app:measurements} Table~\ref{tab:31pgsdssspecresult}.
Two of them need additional notes. For PG1259{+}
593, H$\alpha$ is not covered by the SDSS spectrum because of its high redshift.
In addition, its H$\beta$ region is dominated by the broad component, so the narrow H$\beta$ and [O {\sc iii}] $\lambda5007$ \rev{do not have} reliable measurements. It also has no clear [O~{\sc ii}] $\lambda3727$ emission detected. For PG1415{+}
451, the H$\beta$ region is similarly dominated by the broad component, so the narrow H$\beta$ component is not reliably measured.

For the remaining 56 PG quasars, we first collected publicly available optical spectra from the literature and archival databases. However, many of these spectra are not suitable for our analysis. Some were obtained several decades ago, some have low signal-to-noise ratios or lack associated noise spectra, and some have insufficient wavelength coverage. In particular, after correction to the rest frame, the blue end of many spectra does not cover the [O~{\sc ii}] $\lambda3727$ emission line. We therefore obtained new optical spectroscopic observations for these 56 PG quasars using several optical telescopes, including the Xinglong 2.16 m telescope (XLT), the Lijiang 2.4 m telescope (LJT), the Himalayan Chandra Telescope (HCT), and the Palomar P200/DBSP spectrograph.

In total, we obtain usable optical spectra for 52 sources.
Although the spectra were initially flux-calibrated using spectrophotometric standard stars during the data reduction, we also need additional flux calibration because the observations were obtained with different instruments, under different weather conditions, and at different epochs. Considering the optical photometry used in the SED fitting is based mainly on SDSS photometry, to maintain consistency, we do the flux calibration using SDSS photometric magnitudes.
For a few PG quasars without available SDSS photometry, we instead use Pan-STARRS1 photometry \citep{2016arXiv161205560C,2020ApJS..251....7F}.
The flux-calibrated spectra are then fitted with QSOFITMORE. The key narrow emission lines luminosities and the instrument used for each source are listed in Appendix \ref{app:measurements} Table~\ref{tab:52pgobsspecresult}.
Missing entries are mainly caused by limited wavelength coverage, low spectral resolution, broad-line contamination or low signal-to-noise ratios at the blue end of the spectra, where [O~{\sc ii}] $\lambda3727$ is located.

\subsection{[O~{\sc ii}] Star Formation Rates}
\label{sec:oii_sfr}

\citet{2019ApJ...882...89Z} recalibrated the [O~{\sc ii}] $\lambda3727$ emission line SFR tracer by explicitly including the dependence on gas phase metallicity.
We define the metallicity-dependent term as
\begin{equation}
\label{eq:metallicity_factor}
C(x) =
-4373.14 + 1463.92x - 163.045x^2 + 6.04285x^3 ,
\end{equation}
where $x=12+\log({\rm O/H})$ is the gas-phase oxygen abundance.
For star-forming galaxies, the [O~{\sc ii}] SFR is then given by
\begin{equation}
\label{eq:o2sfr}
{\rm SFR}_{\rm [O\,II]} =
5.3 \times 10^{-42} L_{\rm [O\,II]} C(x)^{-1}.
\end{equation}

For AGNs, \citet{2019ApJ...882...89Z} proposed subtracting the narrow-line region contribution to [O~{\sc ii}] using [O {\sc iii}] $\lambda5007$. The AGN-corrected [O~{\sc ii}] SFR is
\begin{equation}
\label{eq:o2mo3sfr}
{\rm SFR}_{\rm [O\,II]} =
5.3 \times 10^{-42}
\left(L_{\rm [O\,II]} - 0.109L_{\rm [O\,III]}\right)
C(x)^{-1}.
\end{equation}

Here, ${\rm SFR}_{\rm [O\,II]}$ is in units of $M_\odot~{\rm yr}^{-1}$, and $L_{\rm [O\,II]}$ and $L_{\rm [O\,III]}$ are the extinction-corrected line luminosities in units of ${\rm erg~s^{-1}}$. The term $0.109L_{\rm [O\,III]}$ represents the expected [O~{\sc ii}] contribution from the AGN narrow-line region, assuming that [O {\sc iii}] is dominated by AGN photoionization. The remaining [O~{\sc ii}] luminosity is then attributed to star formation in the host galaxy.
The [O~{\sc ii}] SFR calibration adopted from \citet{2019ApJ...882...89Z} is used on the same Kroupa IMF scale as the FIR reference SFRs. This ensures that the comparison between ${\rm SFR}_{\rm [O\,II]}$ and ${\rm SFR}_{\rm FIR}$ is not affected by an IMF-dependent normalization offset.

For quasars with reliable measurements of both narrow H$\alpha$ and narrow H$\beta$, we correct the emission-line luminosities by using the Balmer decrement, adopting an intrinsic line ratio of
$(\mathrm{H}\alpha/\mathrm{H}\beta)_{\rm int}=3.1$ and the \citet{1989ApJ...345..245C} extinction curve with $R_V=3.1$.
For quasars without reliable narrow H$\alpha$ or H$\beta$ measurements or those with a Balmer decrement smaller than 3.1, we instead use the attenuation  from the SED fitting to correct the line luminosities.
We note that this SED extinction correction is used only when the Balmer decrement is unavailable, and it may introduce additional uncertainty because the SED attenuation is less directly tied to the narrow line.
The gas-phase metallicity is estimated from the stellar mass--metallicity relation ($M_\star$--$Z$ relation) \citep{2004ApJ...613..898T,2008ApJ...681.1183K}, using the stellar masses derived from the SED fitting.

We compute the [O~{\sc ii}] SFRs using Equation~(\ref{eq:o2mo3sfr}). For comparison, we also calculate the SFRs using Equation~(\ref{eq:o2sfr}) without subtracting the AGN NLR contribution.
The object-by-object extinction corrections, corrected line luminosities, FIR reference SFRs, AGN-corrected [O~{\sc ii}] SFRs, and [O~{\sc ii}] SFRs without subtracting AGN contribution are provided in Appendix \ref{app:measurements} Table~\ref{tab:o2sfrresult}. A small number of quasars among the 202 objects is not listed because either the [O~{\sc ii}] or [O {\sc iii}] emission-line luminosity cannot be reliably measured.
Therefore, the object numbers in Appendix \ref{app:measurements} Table~\ref{tab:o2sfrresult} are not consecutive.
In the following analysis, we focus on the statistical comparison between the [O~{\sc ii}] SFRs and the FIR reference SFRs.

\subsection{Robustness of the Emission-Line Measurements}
\label{subsec:comparefittingmethod}

\rev{The main analysis models the unresolved [O~{\sc ii}] feature with a single Gaussian and adopts a simplified decomposition for the broad H$\beta$ or H$\alpha$ profile. To assess the influence of these assumptions, we selected 55 SDSS spectra with continuum S/N $>15$ and refitted them with an alternative model. In the alternative fitting, [O~{\sc ii}] was represented by a constrained $\lambda3726,\lambda3729$ doublet with a fixed wavelength separation and a common velocity width. The broad H$\beta$ and H$\alpha$ profiles were modeled with two Gaussian components. The continuum treatment, Fe II template, narrow-line constraints, and treatment of the \linebreak {}[O {\sc iii}] wing were otherwise kept unchanged.}

\rev{Table~\ref{tab:fitting_robustness} compares the results obtained with the two fitting procedures. The integrated [O~{\sc ii}] luminosity is nearly unchanged, with a median difference of $-0.000^{+0.001}_{-0.005}$ dex. Therefore, for this high-quality validation subsample, the single-Gaussian approximation does not introduce a systematic bias in the total [O~{\sc ii}] flux. The alternative broad-H$\beta$ decomposition yields median changes of $-0.12^{+0.08}_{-0.16}$ dex in the narrow H$\beta$ luminosity and $-0.05^{+0.05}_{-0.16}$ dex in the [O {\sc iii}] luminosity.}

\begin{table}[H]

\caption{\rev{{Effect} 
of the alternative emission-line fitting model on the derived quantities for the high-quality validation subsample. The alternative model uses a constrained [O~{\sc ii}] $\lambda3726,\lambda3729$ doublet and two broad H$\beta$ and H$\alpha$ Gaussian components. The quoted values are the median differences, $\Delta Q = Q_{\rm alt}-Q_{\rm orig}$, with the 16th--84th percentile range.}}
\label{tab:fitting_robustness}
\begin{tabularx}{\textwidth}{LCCC}
\toprule
\textbf{Quantity} & \boldmath$N$ & \textbf{Median} \boldmath$\Delta Q$ & \textbf{Unit}\\
\midrule
{$\log L_{\rm [O~\textsc{ii}]}$} 
& 55 & $-0.000^{+0.001}_{-0.005}$ & dex \\
{$\log L_{\rm [O~\textsc{iii}]}$}
& 55 & $-0.05^{+0.05}_{-0.16}$ & dex \\
$\log L_{{\rm H}\beta,\rm narrow}$
& 55 & $-0.12^{+0.08}_{-0.16}$ & dex \\
$\log L_{{\rm H}\alpha,\rm narrow}$
& 29 & $-0.17^{+0.11}_{-0.15}$ & dex \\
$E(B-V)_{\rm Balmer}$
& 29 & $-0.05^{+0.24}_{-0.25}$ & mag \\
{$\log {\rm SFR}_{\rm{[O~\textsc{ii}]-0.109[O~\textsc{iii}]}}$}
& 55 & $-0.08^{+0.38}_{-0.40}$ & dex \\
$\log \lambda_{\rm Edd}$
& 55 & $0.11^{+0.10}_{-0.09}$ & dex \\
\bottomrule
\end{tabularx}

\footnotesize
\rev{{Notes.} 
The $E(B-V)_{\rm Balmer}$ row includes only the sources for which the extinction correction is derived from the narrow-line Balmer decrement. For the remaining sources, the same SED-based attenuation prescription as in the main analysis is retained.
The black-hole masses and Eddington ratios are recalculated using the FWHM of the composite broad H$\beta$ profile, rather than from either individual Gaussian component.}
\end{table}

\rev{For the 29 sources whose extinction correction is based on the narrow-line Balmer decrement, the median change in $E(B-V)$ is $-0.05^{+0.24}_{-0.25}$ mag. After propagation through the extinction correction and the $0.109L_{\rm [O\,III]}$ subtraction, the AGN-corrected [O~{\sc ii}] SFR changes by a median of $-0.08^{+0.38}_{-0.40}$ dex. Thus, the alternative decomposition can affect the derived SFR of individual quasars, but does not introduce a large systematic offset for the validation subsample.
The alternative broad-H$\beta$ model changes the single-epoch black-hole masses by a median of $-0.11^{+0.09}_{-0.10}$ dex and correspondingly increases the Eddington ratios by $0.11^{+0.10}_{-0.09}$ dex. }

\section{Comparison Between [O {\sc \textbf{ii}}] \boldmath$\lambda3727$ and FIR Star Formation Rates}
\label{sec:compareo2firsfr}

\subsection{Comparison Results}

\begin{figure}[H]

\begin{minipage}{0.48\textwidth}

\includegraphics[width=\linewidth]{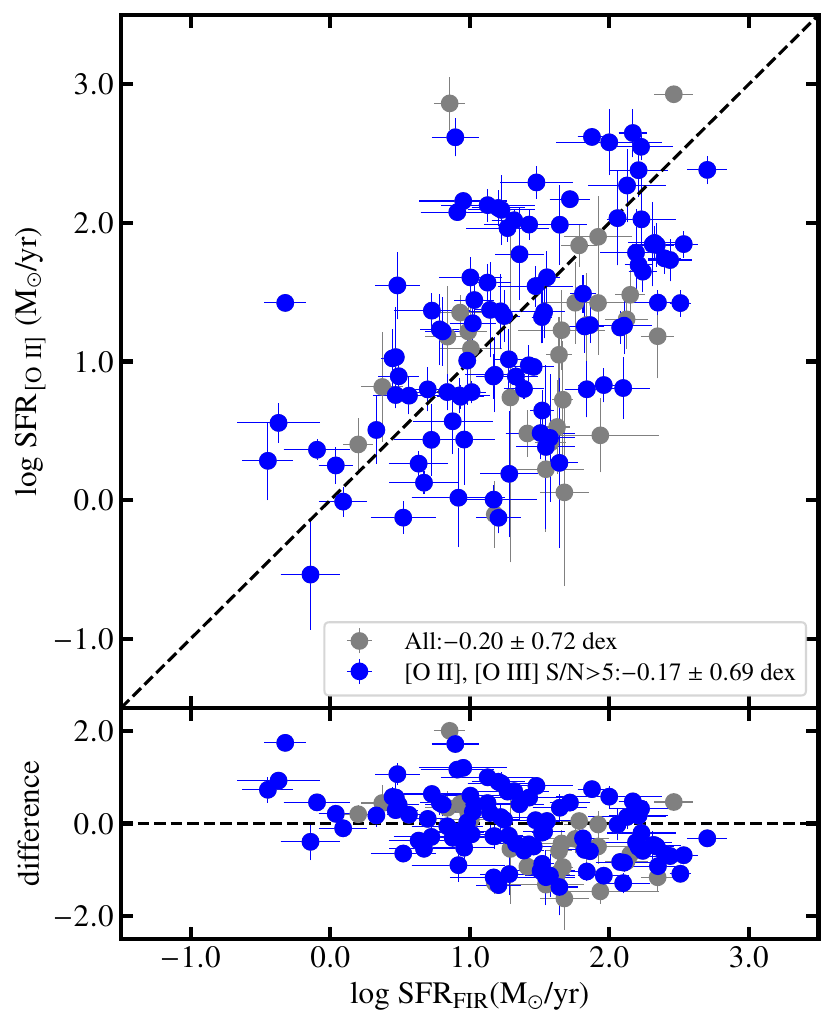}
\centerline{(\textbf{a})}
\end{minipage}
\hfill
\begin{minipage}{0.48\textwidth}

\includegraphics[width=\linewidth]{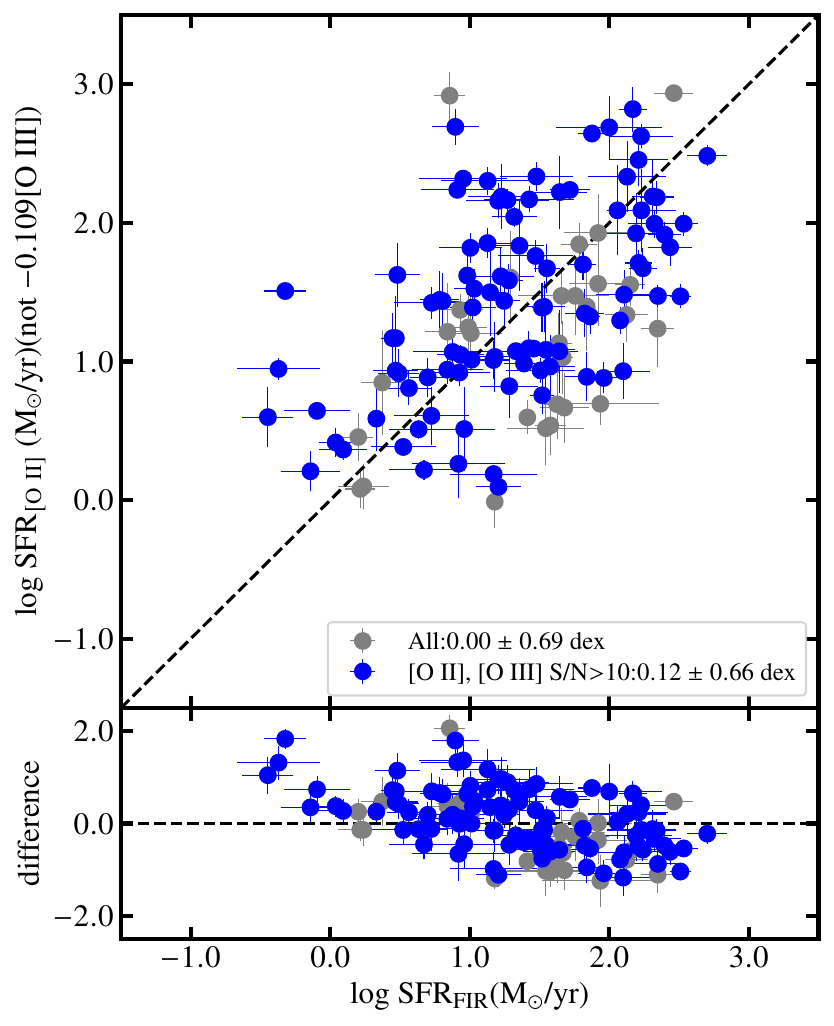}
\centerline{(\textbf{b})}
\end{minipage}
\caption{{Comparison} 
between [O~{\sc ii}] $\lambda3727$ SFRs and FIR SFRs. Panel (\textbf{a}) shows the [O~{\sc ii}] SFR after subtracting the AGN narrow-line contribution estimated as $0.109L_{\rm [O\,III]}$, while panel (\textbf{b}) shows the\linebreak {} [O~{\sc ii}] SFR computed without this correction. In both panels, the lower subpanel shows \mbox{$\log {\rm SFR}_{\rm [O\,II]}-\log {\rm SFR}_{\rm FIR}$}. Blue points indicate sources with ${\rm S/N}>5$ in both [O~{\sc ii}] and [O {\sc iii}]. {The dashed diagonal line in each upper panel marks equality between the [O~{\sc ii}] and FIR SFRs, while the dashed horizontal line in the lower subpanel marks zero difference, $\log {\rm SFR}_{\rm [O\,II]}-\log {\rm SFR}_{\rm FIR}=0$.} The median offset and scatter are labeled in each panel.}
\label{fig:o2sfrsn5}
\end{figure}

We compare the [O~{\sc ii}] $\lambda3727$ SFRs with the FIR SFRs derived from SED fitting in Figure~\ref{fig:o2sfrsn5}.
The abscissa in both panels is the FIR SFR.
In the left panel, the ordinate in the upper subpanel is the [O~{\sc ii}] SFR computed using Equation~(\ref{eq:o2mo3sfr}), where the AGN narrow-line contribution to [O~{\sc ii}] is estimated as $0.109L_{\rm [O\,III]}$ and subtracted from the total \linebreak {}[O~{\sc ii}] luminosity.
In the right panel, the ordinate in the upper subpanel is the [O~{\sc ii}] SFR computed directly from the total [O~{\sc ii}] luminosity using Equation~(\ref{eq:o2sfr}), without subtracting the $0.109L_{\rm [O\,III]}$ term.
In both panels, the lower subpanel shows the difference between the two SFRs, $\log {\rm SFR}_{\rm [O\,II]}-\log {\rm SFR}_{\rm FIR}$. Blue points denote sources with ${\rm S/N}>5$ in both the [O~{\sc ii}] and [O {\sc iii}] emission lines.

As shown in Figure~\ref{fig:o2sfrsn5}a, after subtracting the $0.109L_{\rm [O\,III]}$, the [O~{\sc ii}] SFRs are mainly lower than the FIR SFRs. For the full sample, the median offset is $\log {\rm SFR}_{\rm [O\,II]}-\log {\rm SFR}_{\rm FIR}=-0.20\pm0.72$~dex.
For the subsample with both ${\rm S/N}>5$ [O~{\sc ii}] and [O {\sc iii}], the offset remains similar, $-0.17\pm0.69$~dex.
This suggests that for our quasar sample, the subtracting term $0.109L_{\rm [O\,III]}$ may oversubtract the AGN contribution of [O~{\sc ii}], although the large scatter indicates substantial object-to-object variation.

We therefore also compare the FIR SFRs with the [O~{\sc ii}] SFRs without subtracting $0.109L_{\rm [O\,III]}$, as shown in Figure~\ref{fig:o2sfrsn5}b.
In this case, the median offset becomes $0.00\pm0.69$~dex for the full sample, and the ${\rm S/N}>5$ subsample shows a positive offset of $0.12\pm0.66$~dex.
This indicates that using the total [O~{\sc ii}] luminosity without subtracting AGN narrow line contamination may lead to an overestimation of [O~{\sc ii}] SFR.

One possible explanation is that the [O {\sc iii}] luminosity used for the subtraction is not totally from the AGN NLR. If part of the [O {\sc iii}] emission is from the star-forming regions of host galaxies, then using the total [O {\sc iii}] luminosity to estimate the AGN contribution to [O~{\sc ii}] would oversubtract the [O~{\sc ii}] luminosity.
Another possible explanation is that the coefficient $0.109$ may not be applied to all AGNs, especially quasars. The relative strengths of [O~{\sc ii}] and [O {\sc iii}] in NLR can depend on ionization parameter, gas density, metallicity, and the shape of the ionizing continuum.

In addition, although we use ${\rm SFR}_{\rm FIR}$ as the reference SFR, it can also bring uncertainties.
The FIR SFR are derived from SED decomposition, so they can be affected by different cold dust templates, AGN component, and photometric data coverage, although these uncertainties are smaller than the large scatter between [O~{\sc ii}] SFR and FIR SFR.
In this work, our main goal is to use the comparison to test how well [O~{\sc ii}] performs for the sample as a whole, not to obtain highly precise SFR for each individual quasar.

\begin{figure}[H]

\begin{minipage}{0.48\textwidth}

\includegraphics[width=\linewidth]{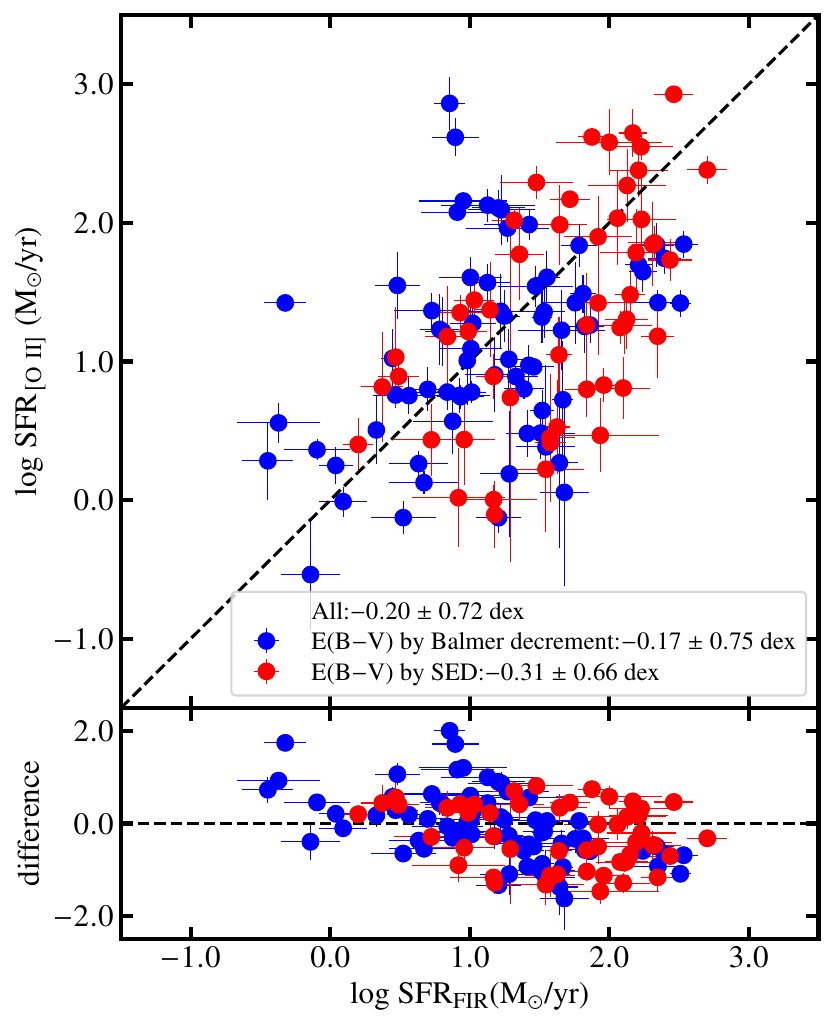}
\centerline{(\textbf{a})}
\end{minipage}
\hfill
\begin{minipage}{0.48\textwidth}

\includegraphics[width=\linewidth]{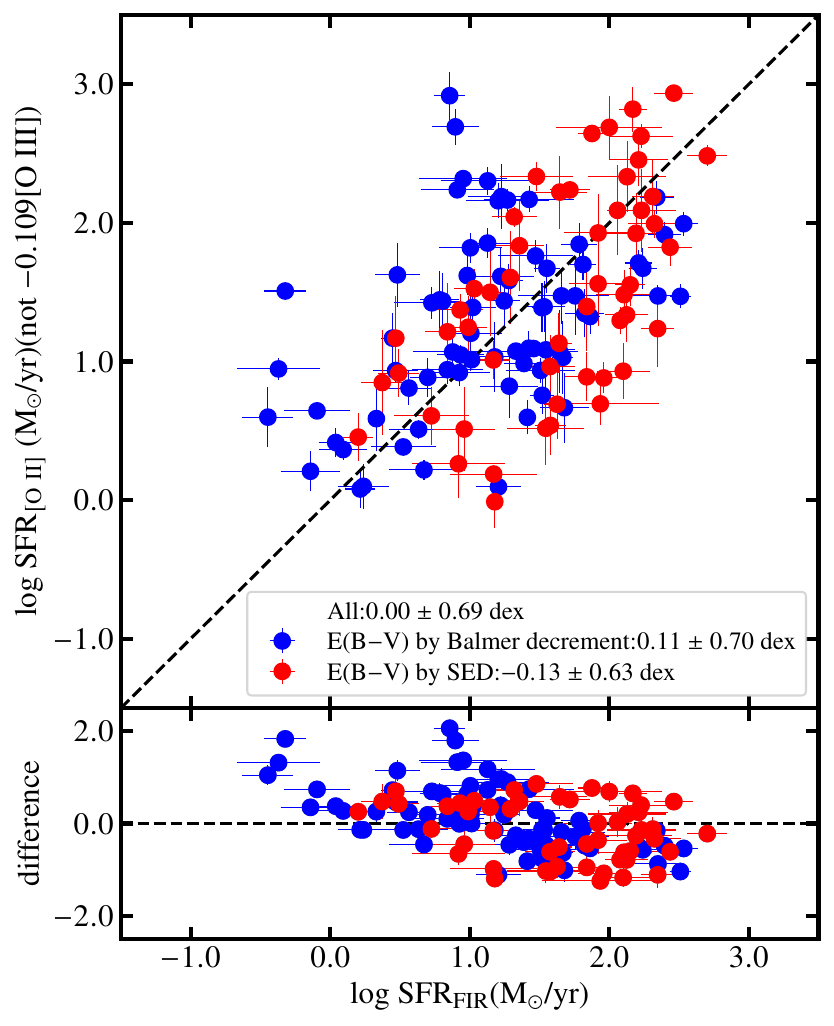}
\centerline{(\textbf{b})}
\end{minipage}
\caption{{Comparison} 
between [O~{\sc ii}] $\lambda3727$ SFRs and FIR SFRs using different extinction-correction methods. The axes {and the dashed reference lines} are the same as in \mbox{Figure~\ref{fig:o2sfrsn5}}. Blue and red points denote sources corrected using the Balmer decrement and SED-fitting attenuation, respectively. Panel (\textbf{a}) shows the [O~{\sc ii}] SFR after subtracting the AGN narrow-line contribution estimated as $0.109L_{\rm [O\,III]}$, while panel (\textbf{b}) shows the \linebreak {}[O~{\sc ii}] SFR computed without this correction.}
\label{fig:o2sfrdiffebmv}
\end{figure}

Systematic uncertainties in the extinction correction may also contribute to the discrepancy.
As described above, we use two methods to correct the [O~{\sc ii}] and [O {\sc iii}] luminosities for dust attenuation: one is based on the Balmer decrement and the other is based on the attenuation from SED fitting.
These two methods may introduce systematic differences in [O~{\sc ii}] SFR results.
We therefore compare the [O~{\sc ii}] SFRs with the FIR SFRs separately for the two extinction correction methods in Figure~\ref{fig:o2sfrdiffebmv}. Blue points represent sources corrected using the Balmer decrement, while red points represent sources corrected using the SED~attenuation.

Figure~\ref{fig:o2sfrdiffebmv} shows that, regardless of whether the $0.109L_{\rm [O\,III]}$ subtraction is applied, the sources corrected using SED attenuation tend to have lower [O~{\sc ii}] SFR than those corrected using the Balmer decrement. This suggests that extinction correction is an important source of systematic uncertainty in the [O~{\sc ii}] SFR.

In addition, the comparison between [O~{\sc ii}] and FIR SFR may be affected by aperture effects. The optical spectra sample the emission-line region within the fiber or slit aperture, whereas the FIR SFR is derived from the integrated dust emission of the host galaxy.
Therefore, spatially extended star formation outside the spectroscopic aperture may cause the [O~{\sc ii}] SFR to be lower than the FIR SFR.
[O~{\sc ii}] and FIR also trace different physical components and timescales, [O~{\sc ii}] traces unobscured or moderately obscured recent star formation in ionized gas, while FIR traces dust-obscured star formation averaged over longer timescales.
These effects, together with metallicity uncertainties and AGN contamination, likely contribute to the large scatter between the two SFR tracers.

Overall, [O~{\sc ii}] can provide a useful but uncertain SFR tracer for quasar host galaxies, and its application requires careful treatment of AGN contamination, extinction correction, metallicity, and aperture effects.

\begin{figure}[H]

\begin{minipage}{0.48\textwidth}

\includegraphics[width=\linewidth]{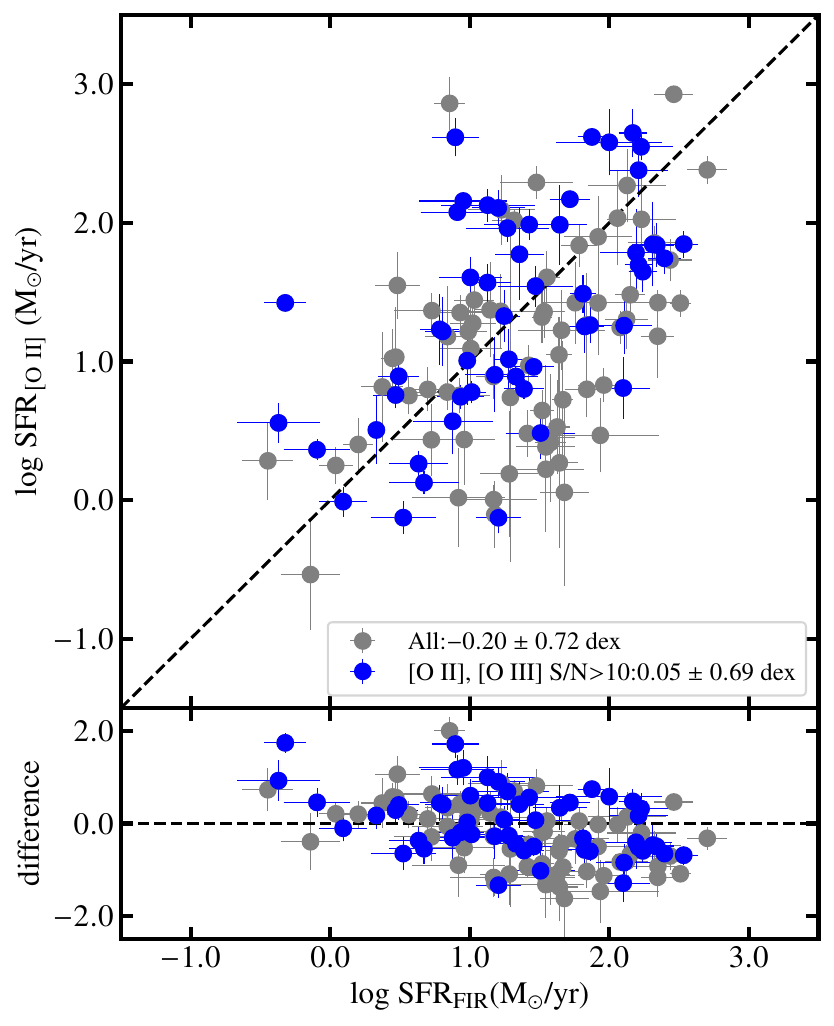}
\centerline{(\textbf{a})}
\end{minipage}
\hfill
\begin{minipage}{0.48\textwidth}

\includegraphics[width=\linewidth]{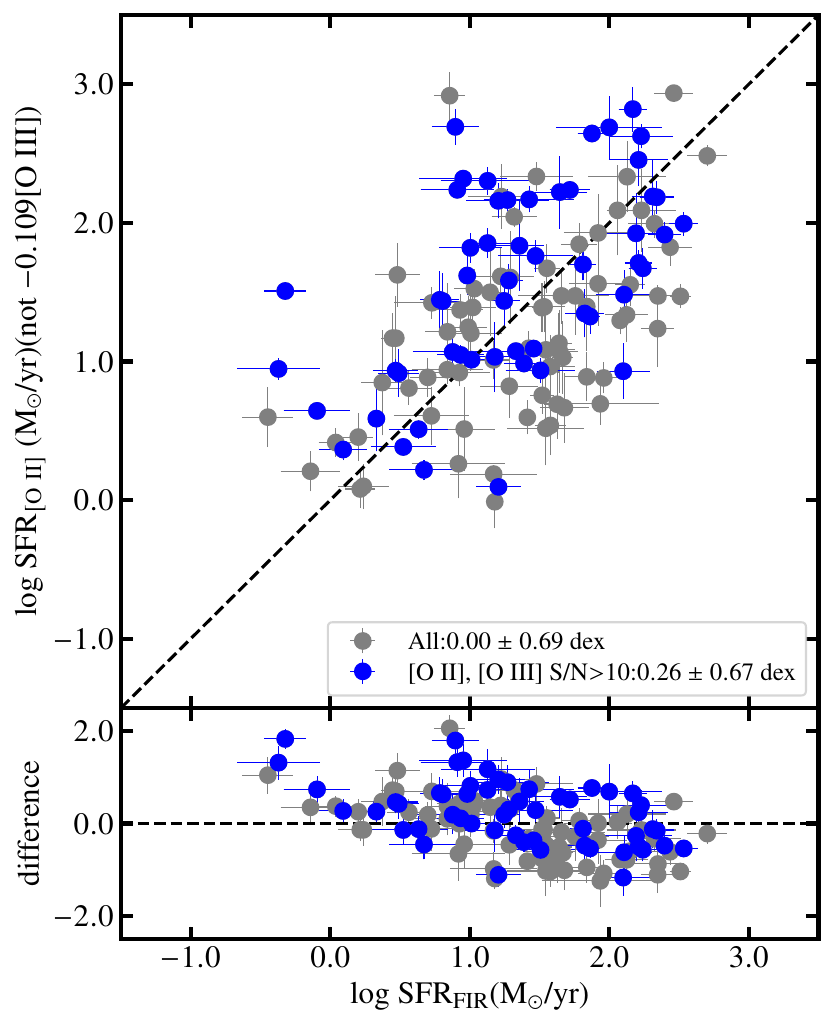}
\centerline{(\textbf{b})}
\end{minipage}
\caption{{Comparison} 
between [O~{\sc ii}] $\lambda3727$ SFRs and FIR SFRs for sources with ${\rm S/N}>10$ in both [O~{\sc ii}] and [O {\sc iii}]. The axes{, symbols and the dashed reference lines} are the same as in Figure~\ref{fig:o2sfrsn5}. Panel (\textbf{a}) shows the [O~{\sc ii}] SFR after subtracting the AGN narrow-line contribution estimated as $0.109L_{\rm [O\,III]}$, while panel (\textbf{b}) shows the [O~{\sc ii}] SFR computed without this correction.}
\label{fig:o2sfrsn10}
\end{figure}

As a further check, since the offset $\log {\rm SFR}_{\rm [O\,II]}-\log {\rm SFR}_{\rm FIR}$ increases when a higher signal-to-noise ratio threshold is applied to the [O~{\sc ii}] and [O {\sc iii}] emission lines, we further raise the threshold to ${\rm S/N}>10$ and examine how the offset changes. The results are shown in Figure~\ref{fig:o2sfrsn10}. For the ${\rm S/N}>10$ subsample, after subtracting the AGN narrow-line contribution estimated as $0.109L_{\rm [O\,III]}$, the median offset is \mbox{$\log {\rm SFR}_{\rm [O\,II]}-\log {\rm SFR}_{\rm FIR}=0.05\pm0.69$~dex}. If this narrow-line correction is not applied, the median offset becomes $0.26\pm0.67$~dex.

For this high-S/N subsample, Figure~\ref{fig:o2sfrsn10}a shows that the [O~{\sc ii}] SFRs corrected using the $0.109L_{\rm [O\,III]}$ term are, on average, comparable to the FIR SFRs.
However, this result should be interpreted with caution. A strict S/N selection prefers to select sources with stronger [O~{\sc ii}] emission, and therefore may bias the sample toward objects with higher [O~{\sc ii}] SFRs. 
Even for this high-S/N subsample, the corrected [O~{\sc ii}] SFRs are only comparable to, rather than significantly higher than, the FIR SFRs, suggesting that the corrected [O~{\sc ii}] SFRs may still be underestimated for some sources.
The right panel shows that, if no correction for the AGN narrow-line contribution is applied, the [O~{\sc ii}] SFRs are systematically higher than the FIR SFRs by 0.26~dex.
This indicates that the AGN contribution to [O~{\sc ii}] is non-negligible and should be accounted for when using [O~{\sc ii}] as an SFR tracer in quasars.
Overall, our comparison suggests that subtracting an AGN narrow-line contribution is necessary, but the simple correction term $0.109L_{\rm [O\,III]}$ may oversubtract the [O~{\sc ii}] luminosity for part of our sample.

Overall, the high-S/N comparison confirms that the AGN contribution to [O~{\sc ii}] cannot be ignored. Without the $0.109L_{\rm [O\,III]}$ subtraction, the [O~{\sc ii}] SFRs are systematically higher than the FIR SFRs. However, the corrected [O~{\sc ii}] SFRs are only comparable to the FIR SFRs even for the high-S/N subsample, suggesting that this correction may oversubtract the star formation-related [O~{\sc ii}] emission in some quasars.

\subsection{Dependence of the SFR Offset on AGN Activity}
\label{subsec:deltasfr_agn}

Since the observed [O~{\sc ii}] $\lambda3727$ emission can originate from both star-forming regions in the host galaxy and the AGN narrow-line region, the difference between the [O~{\sc ii}] and FIR SFRs may depend on AGN activity. We therefore compare the SFR offset,
\begin{equation}
\Delta \log {\rm SFR} =
\log {\rm SFR}_{\rm [O\,II]} - \log {\rm SFR}_{\rm FIR},
\end{equation}
with the bolometric luminosity, $L_{\rm bol}$, and the Eddington ratio, $\lambda_{\rm Edd}=L_{\rm bol}/L_{\rm Edd}$. The Eddington luminosity is given by
\begin{equation}
L_{\rm Edd} =
1.26 \times 10^{38}
\left(\frac{M_{\rm BH}}{M_\odot}\right)
~{\rm erg~s^{-1}} .
\end{equation}

\begin{figure}[H]

\begin{minipage}{0.48\textwidth}

\includegraphics[width=\linewidth]{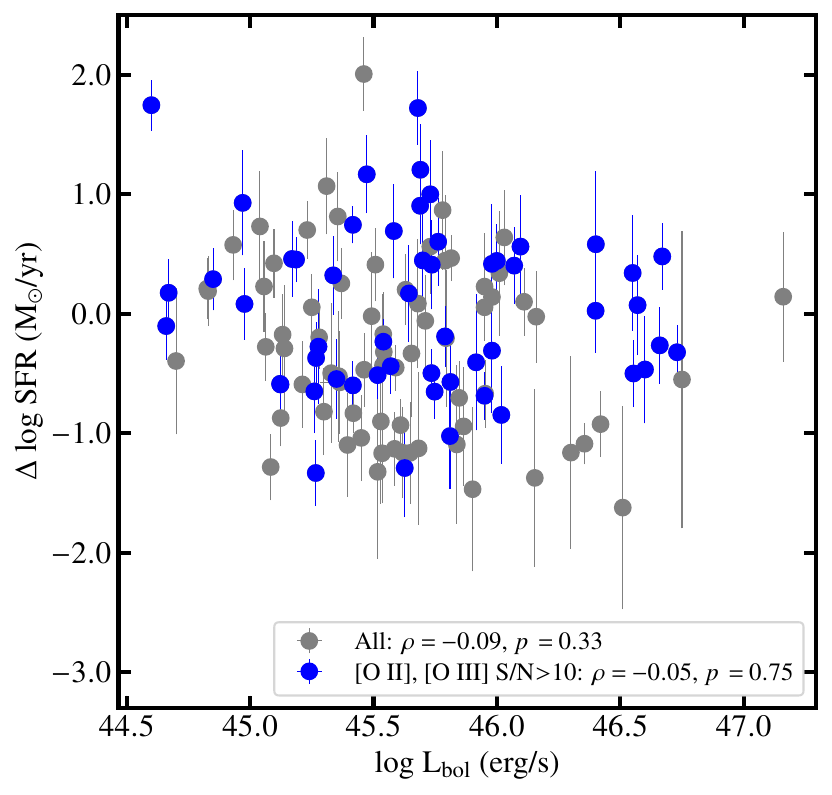}
\centerline{(\textbf{a})}
\end{minipage}
\hfill
\begin{minipage}{0.48\textwidth}

\includegraphics[width=\linewidth]{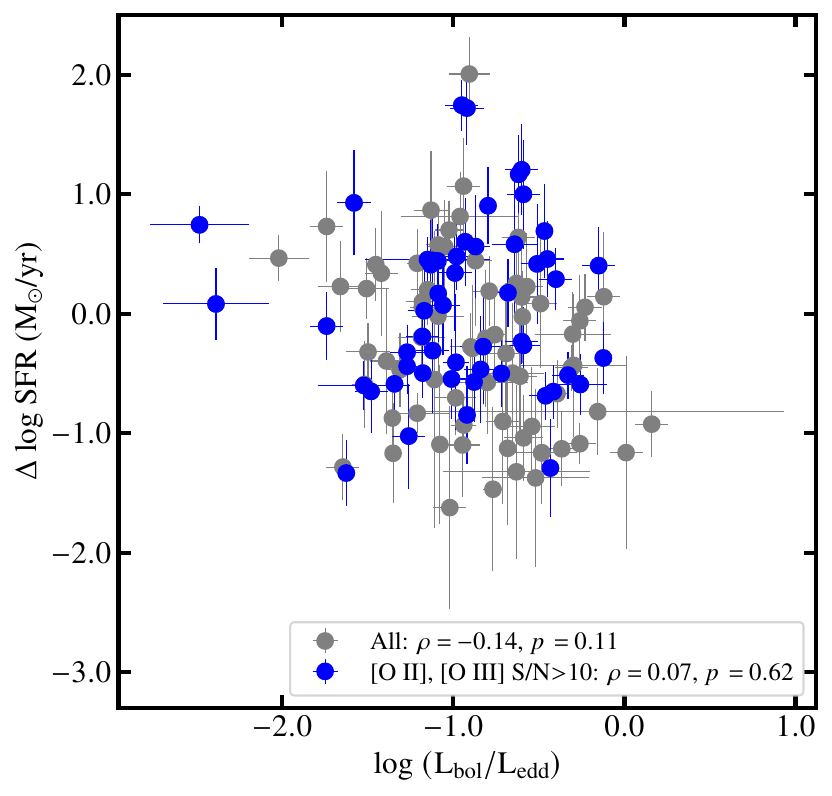}
\centerline{(\textbf{b})}
\end{minipage}
\caption{{SFR} 
offset, $\Delta \log {\rm SFR} = \log {\rm SFR}_{\rm [O\,II]} - \log {\rm SFR}_{\rm FIR}$, as a function of (\textbf{a}) bolometric luminosity and (\textbf{b}) Eddington ratio. The Spearman correlation coefficient, $\rho$, and the corresponding $p$ value are shown in the lower right corner of each panel for both the full sample and the subsample with ${\rm S/N}>10$ in both [O~{\sc ii}] and [O {\sc iii}].}
\label{fig:lboleddratiodeltasfr}
\end{figure}

The results are shown in Figure~\ref{fig:lboleddratiodeltasfr}. The left panel compares $\Delta \log {\rm SFR}$ with $L_{\rm bol}$, and the right panel compares $\Delta \log {\rm SFR}$ with $\lambda_{\rm Edd}$.
The Spearman correlation coefficient, $\rho$, and the corresponding $p$ value are shown in the lower right corner of each panel.
For both the full sample and the subsample with ${\rm S/N}>10$ [O~{\sc ii}] and [O {\sc iii}], there is no statistically significant correlation between $\Delta \log {\rm SFR}$ and $L_{\rm bol}$, although a weak negative trend is visually suggested. For $\lambda_{\rm Edd}$, the full sample shows only a very weak negative trend, and no significant correlation is found for the high S/N subsample.

\begin{figure}[H]

\includegraphics[width=0.65\textwidth]{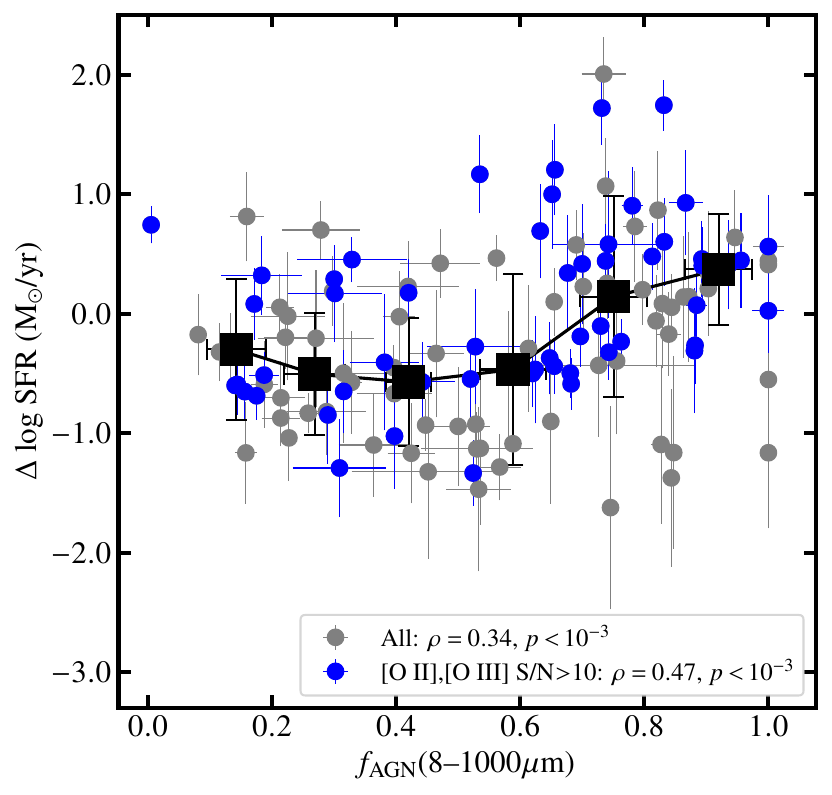}
\caption{{SFR} 
offset, $\Delta \log {\rm SFR} = \log {\rm SFR}_{\rm [O\,II]} - \log {\rm SFR}_{\rm FIR}$, as a function of the fractional AGN contribution to the total infrared luminosity, $f_{\rm AGN}(8$--$1000~\upmu\mathrm{m})$. The Spearman correlation coefficient, $\rho$, and the corresponding $p$ value are shown in the lower right corner for both the full sample and the subsample with ${\rm S/N}>10$ in both [O~{\sc ii}] and [O {\sc iii}]. Black squares show the median values in \mbox{six bins} of $f_{\rm AGN}$.}
\label{fig:fracagndeltasfr}
\end{figure}

\textls[-15]{We also compare $\Delta \log {\rm SFR}$ with the fractional AGN contribution to the total infrared luminosity, $f_{\rm AGN}$ (8--1000~\textmu m), derived from the SED fitting. The result is shown in Figure~\ref{fig:fracagndeltasfr}. For the full sample, the Spearman correlation coefficient is $\rho=0.34$, with $p<10^{-3}$,} indicating a weak positive correlation. For the high-S/N subsample, the correlation becomes somewhat stronger, with $\rho=0.47$ and $p<10^{-3}$. To illustrate the overall trend, we divide the sample into six bins in $f_{\rm AGN}$ (8--1000~\textmu m) and show the median value in each bin as black squares.

For sources with $f_{\rm AGN}$ (8--1000~\textmu m) $\lesssim$ 0.6, $\Delta \log {\rm SFR}$ shows no clear dependence on $f_{\rm AGN}$, and the [O~{\sc ii}] SFRs tend to be lower than the FIR SFRs. At higher $f_{\rm AGN}$, however, $\Delta \log {\rm SFR}$ increases with increasing $f_{\rm AGN}$ and can become positive, indicating that the [O~{\sc ii}] SFRs exceed the FIR SFRs for sources with large AGN infrared contributions.
This behavior may have several possible origins. First, a larger $f_{\rm AGN}$ (8--1000~\textmu m) implies a smaller relative contribution from the cold dust component used to derive ${\rm SFR}_{\rm FIR}$, which can \textls[-15]{lower the FIR SFR.
Second, when the AGN contribution is weak, a non-negligible fraction of the observed [O {\sc iii}] emission may be associated with star-forming regions in the host galaxy. In this case, using the total $L_{\rm [O\,III]}$ to estimate the AGN narrow-line contribution to [O~{\sc ii}] may over-subtract the star formation-related [O~{\sc ii}] emission, leading to underestimated [O~{\sc ii}] SFRs.}

However, the correlation with $f_{\rm AGN}$ should be interpreted with caution. Both $f_{\rm AGN}$ and ${\rm SFR}_{\rm FIR}$ are derived from the same SED decomposition, so part of the observed trend may reflect covariance between the AGN and cold-dust components in the SED fitting. In addition, uncertainties in dust attenuation, metallicity, aperture effects, and the different timescales traced by optical emission lines and FIR emission may also contribute to the scatter in $\Delta \log {\rm SFR}$.

\subsection{Dependence of the SFR Offset on Redshift}
\label{subsec:deltasfr_redshift}

\begin{figure}[H]

\includegraphics[width=0.65\textwidth]{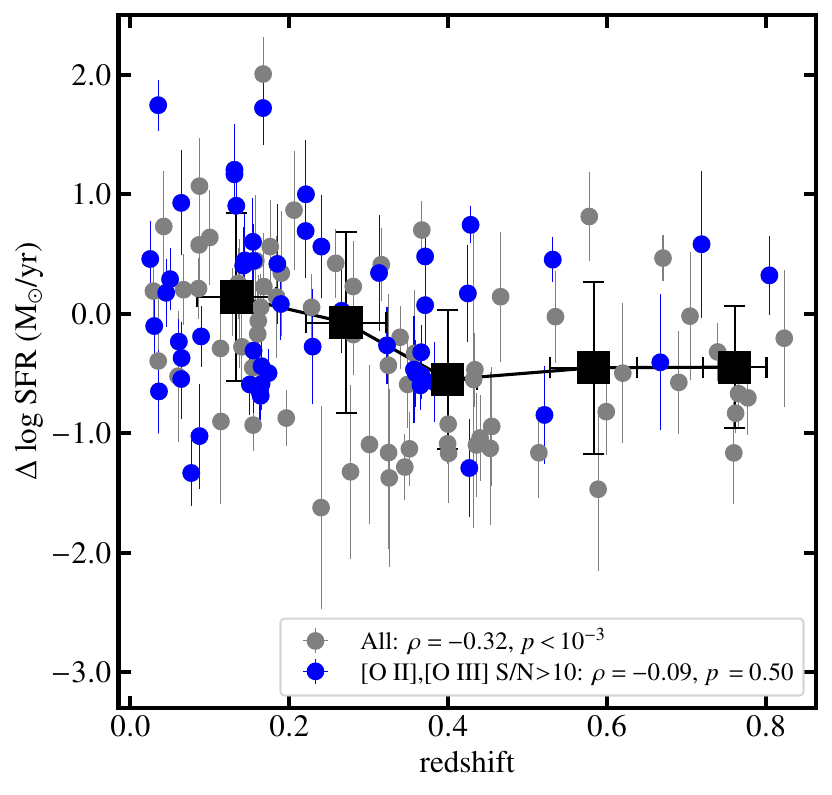}
\caption{{SFR} 
offset, $\Delta \log {\rm SFR}$, as a function of redshift. The Spearman correlation coefficient, $\rho$, and the corresponding $p$ value are shown in the lower right corner for both the full sample and the subsample with ${\rm S/N}>10$ in both [O~{\sc ii}] and [O {\sc iii}]. Black squares show the median values in \mbox{five redshift} bins.}
\label{fig:zdeltasfr}
\end{figure}

We further examine whether the SFR offset ($\Delta \log {\rm SFR}$) depends on redshift. The result is shown in Figure \ref{fig:zdeltasfr}. For the full sample, the Spearman correlation coefficient is $\rho=-0.32$, with $p<10^{-3}$, indicating a weak negative correlation between $\Delta \log {\rm SFR}$ and redshift. To better illustrate the overall trend, we divide the sample into five redshift bins and show the median value in each bin as black squares.

For quasars at $z<0.4$, $\Delta \log {\rm SFR}$ decreases gradually with increasing redshift. At $z>0.4$, the median offset becomes relatively flat, but the [O~{\sc ii}] SFRs are generally lower than the FIR SFRs. In particular, the offset is close to zero for the low-redshift subsample at $z<0.2$, while the [O~{\sc ii}] SFRs tend to be underestimated relative to the FIR SFRs for sources at higher redshift. This suggests that, for quasars at $z\gtrsim0.4$--0.5, the [O~{\sc ii}] SFRs derived using Equation~(\ref{eq:o2mo3sfr}) should be interpreted with caution.
This redshift dependence may partly arise from methodological effects, including the lack of H$\alpha$ coverage at higher redshift, increased reliance on SED attenuation corrections, aperture effects, FIR selection bias, and uncertainties in the AGN NLR correction, rather than pure physical evolution.

\subsection{\rev{Sensitivity to the Adopted AGN NLR Correction}}
\label{subsec:comparenlrcorrection}

\rev{The comparisons above show that neither ignoring the AGN contribution nor applying a fixed $0.109L_{\rm [O\,III]}$ subtraction can provide an equally accurate [O~{\sc ii}] SFR for every quasar. If no AGN correction is applied, the observed [O~{\sc ii}] emission may still include a contribution from the AGN NLR. On the other hand, using the same $0.109L_{\rm [O\,III]}$ correction for all sources may remove too much or too little AGN-related [O~{\sc ii}] emission in some objects. We therefore compare the SFR offset ($\Delta \log {\rm SFR} = \log {\rm SFR}_{\rm [O\,II]} - \log {\rm SFR}_{\rm FIR}$)
calculated with and without subtracting $0.109L_{\rm [O\,III]}$. We examine how this offset changes with AGN luminosity, Eddington ratio, infrared AGN fraction, and redshift.}

\begin{figure}[H]

\begin{minipage}[b]{0.495\textwidth}
\centering
\includegraphics[width=\linewidth]{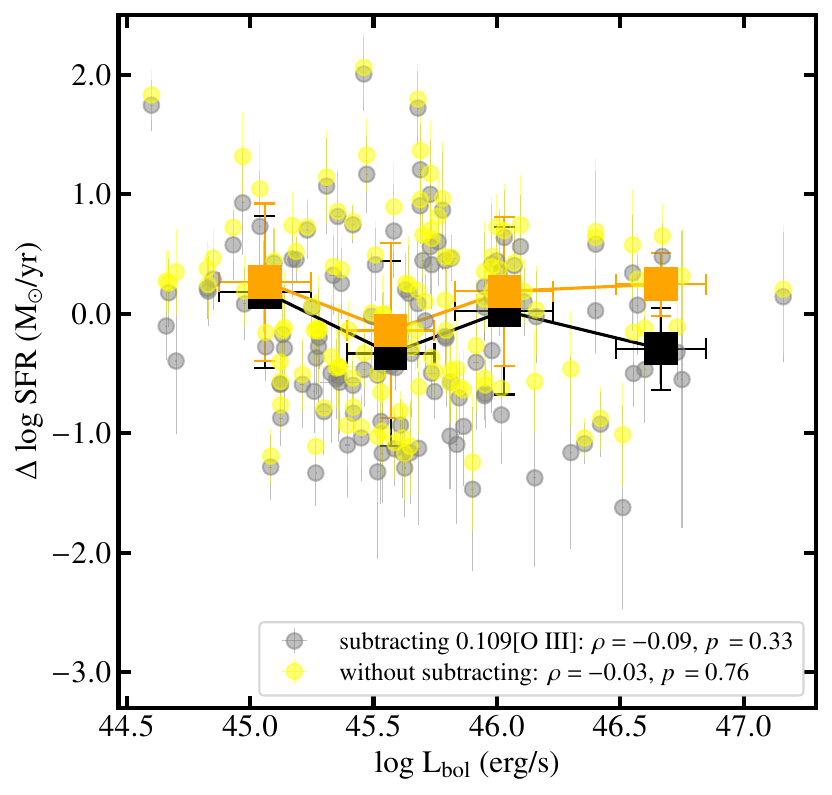}
\par\smallskip
(\textbf{a}) $L_{\rm bol}$
\end{minipage}
\hfill
\begin{minipage}[b]{0.495\textwidth}
\centering
\includegraphics[width=\linewidth]{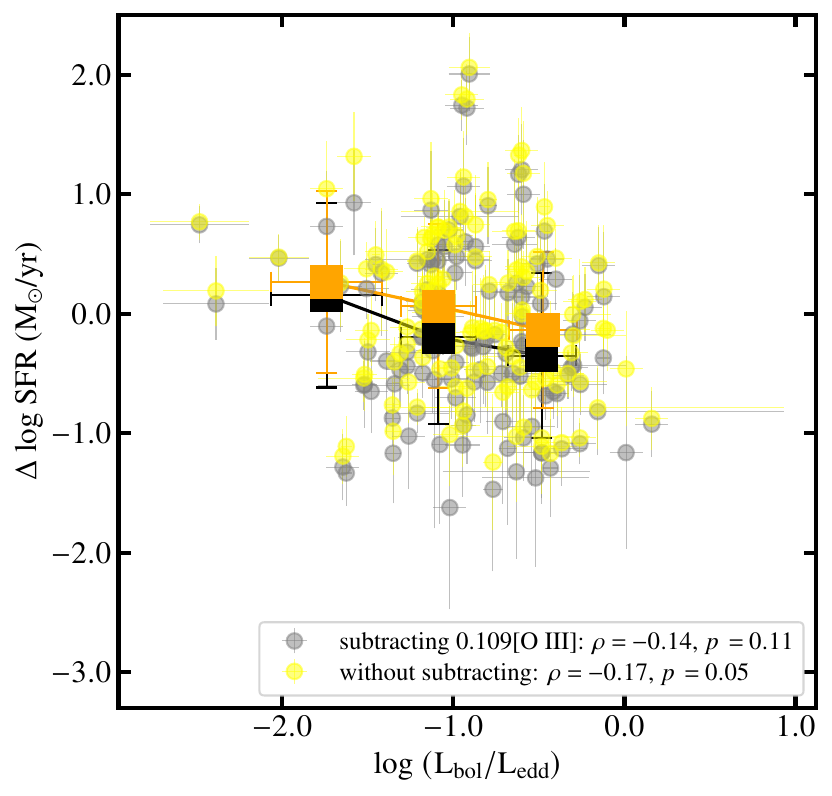}
\par\smallskip
(\textbf{b}) $\lambda_{\rm Edd}$
\end{minipage}

\medskip

\begin{minipage}[b]{0.495\textwidth}
\centering
\includegraphics[width=\linewidth]{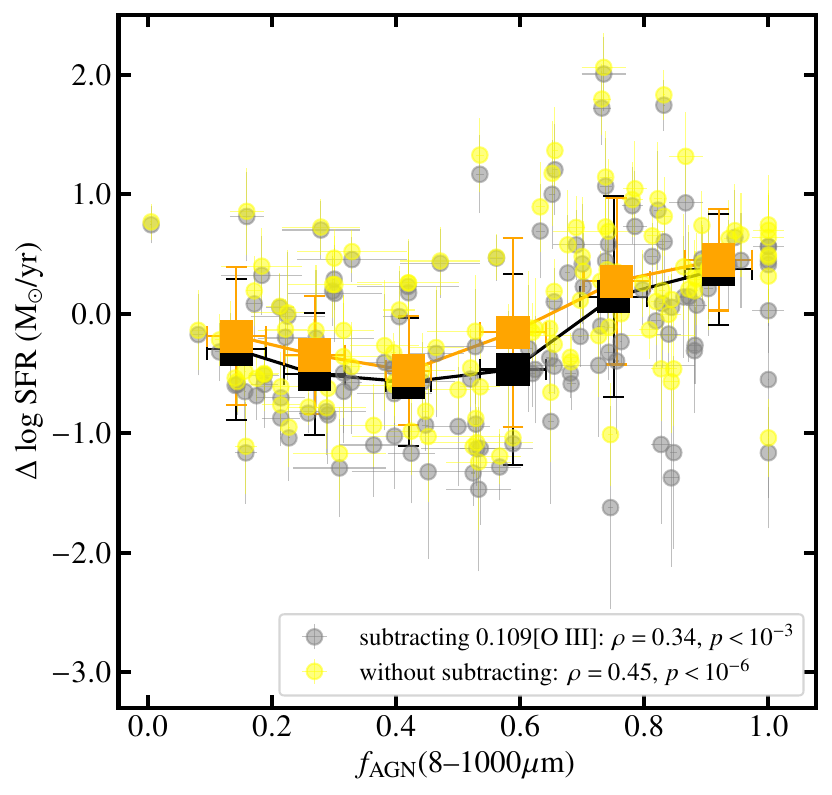}
\par\smallskip
(\textbf{c}) $f_{\rm AGN}$ (8--1000~\textmu m)
\end{minipage}
\hfill
\begin{minipage}[b]{0.495\textwidth}
\centering
\includegraphics[width=\linewidth]{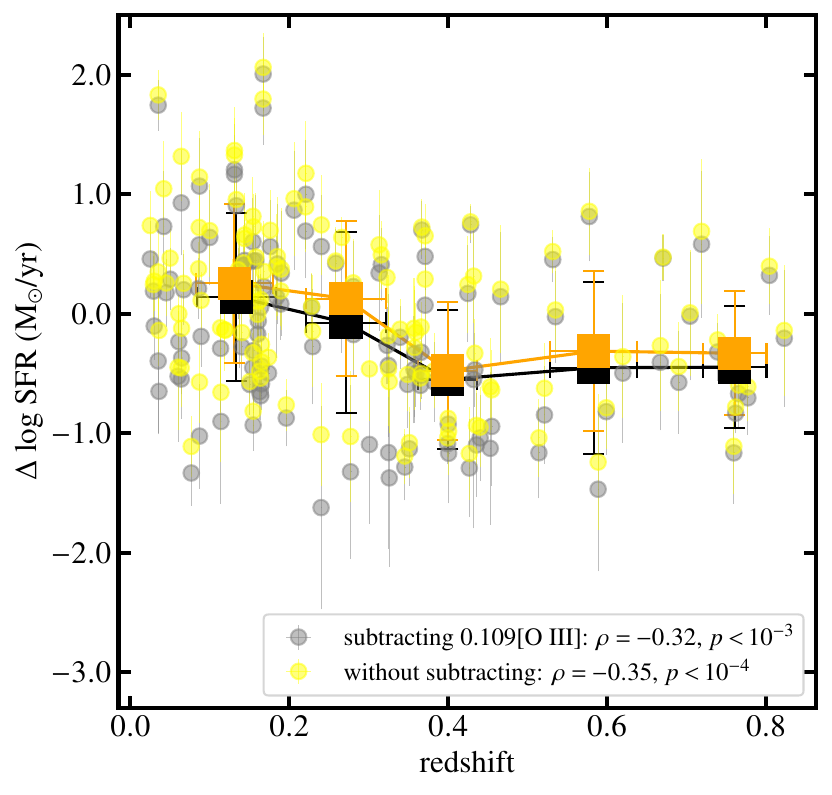}
\par\smallskip
(\textbf{d}) Redshift
\end{minipage}

\caption{\rev{{Comparison} 
of the SFR offset, $\Delta \log {\rm SFR}=\log {\rm SFR}_{\rm [O\,II]}-\log {\rm SFR}_{\rm FIR}$, derived with and without subtracting the AGN narrow-line contribution $0.109L_{\rm [O\,III]}$. The panels show the dependence on (\textbf{a}) bolometric luminosity, (\textbf{b}) Eddington ratio, (\textbf{c}) infrared AGN fraction and (\textbf{d}) redshift. The two sets of symbols correspond to the results with and without the $0.109L_{\rm [O\,III]}$ subtraction, as indicated in each panel. {Black and orange squares show the median values in bins for the results with and without the $0.109L_{\rm [O~{\,III}]}$ subtraction, respectively.} 
}}
\label{fig:agn_correction_comparison}
\end{figure}

\rev{Figure~\ref{fig:agn_correction_comparison} shows no statistically significant dependence of the SFR offset on $L_{\rm bol}$ for either treatment. The corresponding Spearman coefficients are $\rho=-0.09$ ($p=0.33$) after the subtraction and $\rho=-0.03$ ($p=0.76$) without the subtraction. Similarly, the dependence on $\lambda_{\rm Edd}$ is weak. The uncorrected result shows a little negative correlation ($\rho=-0.17$, $p=0.05$), while the corrected result is not statistically significant ($\rho=-0.14$, $p=0.11$). So, within the dynamic range of our sample, we find no strong evidence that the relative performance of the fixed correction depends systematically on AGN luminosity or accretion state.
As for the infrared AGN fraction, the SFR offset increases with $f_{\rm AGN}$ both before and after the subtraction, but the correlation becomes weaker after the $0.109L_{\rm [O\,III]}$ term is applied, decreasing from $\rho=0.45$ ($p<10^{-6}$) to $\rho=0.34$ ($p<10^{-3}$). This suggests that the adopted correction removes part of the AGN-related excess in the observed [O~{\sc ii}] emission. However, the remaining positive trend indicates that a single [O~{\sc ii}]/[O {\sc iii}] ratio does not fully describe the object-to-object diversity of the quasar NLR. However, this trend should be interpreted with caution because $f_{\rm AGN}$ and ${\rm SFR}_{\rm FIR}$ are both derived from the same SED decomposition and are therefore not completely independent.
The SFR offset decreases with redshift both with and without the AGN correction, with $\rho=-0.32$ ($p<10^{-3}$) and $\rho=-0.35$ ($p<10^{-4}$), respectively. Since the redshift trend is present with comparable strength in both cases, it is unlikely to be caused primarily by the fixed $0.109L_{\rm [O,III]}$ subtraction. Instead, it may reflect a combination of aperture effects, changes in the extinction-correction method when narrow H$\alpha$ is unavailable, sample selection, and the different physical timescales traced by optical emission lines and FIR emission.}

\rev{Overall, the comparison indicates that the AGN contribution to [O~{\sc ii}] cannot be ignored statistically, but that the fixed $0.109L_{\rm [O,III]}$ subtraction should not be interpreted as an exact object-by-object correction. It reduces part of the AGN-related bias, particularly for sources with a large infrared AGN contribution, while substantial residual scatter remains.}

\rev{A recent study by \citet{2026ApJ..1001L..15W} used a kinematically consistent decomposition of [O~{\sc ii}] $\lambda3727$ and the high-ionization [Ne {\sc v}] $\lambda3426$ line to estimate the AGN contribution to [O~{\sc ii}] emission in SDSS quasars. They found that the narrow [O~{\sc ii}] emission is dominated by star formation on average, with an estimated AGN contribution of only $\sim$6\% in their composite spectrum. When weak broad and very broad [O~{\sc ii}] components were included, the inferred AGN contribution increased to $\sim$17\%. Their result supports the use of [O~{\sc ii}] as a statistical SFR tracer in quasars. However, they also noted that the AGN contribution can vary among individual sources and that reliable measurements of the [O~{\sc ii}] and [Ne {\sc v}] line profiles are needed for an object-by-object correction.
Our results are broadly consistent with this picture. The comparison with FIR SFRs indicates that AGN contamination in [O~{\sc ii}] cannot be ignored, especially for sources with strong AGN-related emission. At the same time, the systematic offset introduced by the fixed $0.109L_{\rm [O\,III]}$ subtraction suggests that a universal [O~{\sc ii}]/[O {\sc iii}] ratio may not provide an accurate correction for every quasar. A line-profile-based method using [Ne {\sc v}], such as that of \citet{2026ApJ..1001L..15W}, may provide a useful alternative when spectra of sufficient quality are available.}

\section{Summary and Conclusions}
\label{sec:summary}

\rev{We} select a sample of 202 quasars, including 120 SDSS quasars and 87 PG quasars, with five overlapping objects. We measure [O~{\sc ii}] $\lambda3727$, [O {\sc iii}] $\lambda5007$, and narrow Balmer emission lines by fitting SDSS spectra and newly observed optical spectra of PG quasars.
Then, we investigate the reliability of the [O~{\sc ii}] $\lambda3727$ emission line as an SFR tracer for the quasar host galaxies by comparing [O~{\sc ii}] SFRs with FIR SFRs.
The FIR SFRs are derived from multiwavelength SED decomposition and are used as an independent SFR reference. Our main results are summarized as follows.

\begin{enumerate}

\item We refit the SDSS spectra of 120 SDSS quasars and 31 PG quasars with available SDSS spectra using QSOFITMORE.
For the remaining PG quasars without suitable archival spectra, we obtained new optical spectroscopic observations with XLT, LJT, HCT, and P200/DBSP. We got new spectra for 52 PG quasars. These spectra-fitting results provide the narrow emission line luminosities for calculating [O~{\sc ii}] SFRs.

\item We calculate [O~{\sc ii}] SFRs using the metallicity-dependent calibration of \citet{2019ApJ...882...89Z}. We subtract the AGN NLR contribution to [O~{\sc ii}], estimated as $0.109L_{\rm [O\,III]}$. Extinction is corrected using the Balmer decrement when reliable narrow H$\alpha$ and H$\beta$ measurements are available. Otherwise, we use the extinction obtained from SED fitting. The gas metallicity is estimated from the stellar mass--metallicity relation.

\item After subtracting the $0.109L_{\rm [O\,III]}$ contribution, [O~{\sc ii}] SFRs are mainly lower than FIR SFRs. The median offset is $\log{\rm SFR}_{\rm [O\,II]}-\log{\rm SFR}_{\rm FIR}=-0.20\pm0.72$~dex for the full sample and $-0.17\pm0.69$~dex for sources with ${\rm S/N}>5$ [O~{\sc ii}] and [O {\sc iii}].
This suggests that the $0.109L_{\rm [O\,III]}$ correction may oversubtract the star formation-related [O~{\sc ii}] luminosity for some quasars, although the large scatter indicates substantial object-to-object variation.

\item If the AGN NLR subtraction is not applied, the median offset becomes $0.00\pm0.69$~dex for the full sample and $0.12\pm0.66$~dex for the ${\rm S/N}>5$ subsample. For the more strict ${\rm S/N}>10$ subsample, the corrected [O~{\sc ii}] SFRs are comparable to FIR SFRs, with a median offset of $0.05\pm0.69$~dex, while the uncorrected [O~{\sc ii}] SFRs are higher by $0.26\pm0.67$~dex. These results show that the AGN contribution to [O~{\sc ii}] is non-negligible and should be considered when using [O~{\sc ii}] as an SFR tracer of quasar host~galaxies.

\item The extinction correction is an important source of systematic uncertainty. Sources corrected using the SED extinction tend to have lower [O~{\sc ii}] SFRs than sources corrected using the Balmer decrement. This indicates that the uncertainty of the narrow-line gas extinction can contribute significantly to the scatter between [O~{\sc ii}] and FIR SFRs.

\item We find no statistically significant correlation between the SFR offset and bolometric luminosity or Eddington ratio. However, the SFR offset shows a weak positive correlation with the fractional AGN contribution to the total infrared luminosity, $f_{\rm AGN}$ (8--1000 \textmu m), with $\rho=0.34$ and $p<10^{-3}$ for the full sample. The correlation is stronger for the high S/N subsample. This trend may reflect the combined effects of AGN contamination, SED decomposition component degeneracy, and the relative contributions of AGN-heated dust and cold dust.

\item The SFR offset shows a weak negative correlation with redshift for the full sample, with $\rho=-0.32$ and $p<10^{-3}$. The [O~{\sc ii}] SFRs are close to the FIR SFRs at $z<0.2$, but tend to be lower than the FIR SFRs at $z\gtrsim0.4$--0.5. This redshift dependence may partly arise from methodological effects, including the lack of H$\alpha$ coverage at higher redshift, increased reliance on SED extinction corrections, aperture effects, FIR selection bias, and uncertainties in the AGN NLR correction.

\end{enumerate}

Overall, our results show that [O~{\sc ii}] $\lambda3727$ is useful for studying the statistical SFR properties of quasar host galaxies, but it has large uncertainties for individual objects.
An AGN NLR correction is needed to reduce AGN contamination. Otherwise, the [O~{\sc ii}] SFR may be overestimated.
In addition, a fixed subtraction of $0.109L_{\rm [O\,III]}$ may oversubtract the [O~{\sc ii}] luminosity of some quasars.
Therefore, reliable use of [O~{\sc ii}] requires careful consideration of AGN contamination, dust extinction, metallicity, aperture effects, and redshift-related biases.

\vspace{+6pt}
\authorcontributions{Conceptualization, X.F. and X.-B.W.; methodology, X.F., X.-B.W. and Y.F.; software, X.F. and Y.F.; validation, X.F. and Y.F.; formal analysis, X.F.; investigation, X.F., X.-B.W., Y.F., Y.P., R.Z. and H.W.; resources, X.F., X.-B.W., Y.F. and Y.P.; data curation, X.F., X.-B.W., Y.F., Y.P., R.Z. and H.W.; writing---original draft preparation, X.F.; writing---review and editing, X.F., X.-B.W., Y.F., Y.P., R.Z. and H.W.; visualization, X.F.; supervision, X.-B.W.; project administration, X.-B.W.; funding acquisition, X.-B.W. All authors have read and agreed to the published version of the manuscript.}

\funding{This research was funded by the National \rev{Key} R\&D Program of China, grant numbers 2025YFA1614100 and 2025YFA1614101, and the National Science Foundation of China, grant \mbox{number 12133001.}}

\dataavailability{The public data used in this study are available from the Sloan Digital Sky Survey, the NASA/IPAC Infrared Science Archive, the NASA/IPAC Extragalactic Database, the Herschel Science Archive, and Pan-STARRS1. The newly obtained optical spectra and derived measurements presented in this work are available from the corresponding author upon reasonable~request.}

\acknowledgments{We acknowledge the support of the staff of the Xinglong 2.16 m telescope and the Lijiang 2.4 m telescope, which was partially supported by the Open Project Program of the Key Laboratory of Optical Astronomy, National Astronomical Observatories, Chinese Academy of Sciences and funded by the Chinese Academy of Sciences and the People's Government of Yunnan Province, respectively. The authors also thank the staff of the Himalayan Chandra Telescope and the Palomar Observatory for their support of the optical spectroscopic observations. This work makes use of SDSS spectroscopic data and public photometric data from the NASA/IPAC Infrared Science Archive (IRSA). This research has also made use of the {NASA/IPAC Extragalactic Database (NED),} 
which is funded by the National Aeronautics and Space Administration and operated by the California Institute of Technology. This work is based in part on data from the Herschel Space Observatory, an ESA space observatory with science instruments provided by European-led Principal Investigator consortia and with important participation from NASA. This work also makes use of data products from Pan-STARRS1. {Funding for the Sloan Digital Sky Survey has been provided by the Alfred P. Sloan Foundation, the Participating Institutions, the National Science Foundation, the U.S. Department of Energy, the National Aeronautics and Space Administration, the Japanese Monbukagakusho, the Max Planck Society, and the Higher Education Funding Council for England.} 
}

\conflictsofinterest{The authors declare no conflicts of interest.}


\newpage

\abbreviations{Abbreviations}{
The following abbreviations are used in this manuscript:
\\

\noindent
\begin{tabular}{@{}ll}
AGN & Active galactic nucleus \\
BC & Balmer continuum \\
DBSP & Double Spectrograph \\
FIR & Far-infrared \\
FWHM & Full width at half maximum \\
HCT & Himalayan Chandra Telescope \\
IRSA & NASA/IPAC Infrared Science Archive \\
JCMT & James Clerk Maxwell Telescope \\
LJT & Lijiang 2.4 m Telescope \\
NED & NASA/IPAC Extragalactic Database \\
NLR & Narrow-line region \\
PACS & Photodetector Array Camera and Spectrometer \\
PCA & Principal component analysis \\
PG & Palomar-Green \\
SDSS & Sloan Digital Sky Survey \\
SED & Spectral energy distribution \\
SFR & Star formation rate \\
S/N & Signal-to-noise ratio \\
SPIRE & Spectral and Photometric Imaging Receiver \\
XLT & Xinglong 2.16 m Telescope \\
\end{tabular}
}

\appendixtitles{yes}
\appendixstart
\appendix

\section{Emission-Line and Star Formation Rate Measurements}
\label{app:measurements}

This appendix provides the object-by-object measurements used in the analysis. \mbox{Appendix} Tables~\ref{tab:115sdssspecresult}--\ref{tab:52pgobsspecresult} list the narrow emission-line luminosities measured from the SDSS and newly obtained optical spectra.
Appendix \ref{app:measurements} Table~\ref{tab:o2sfrresult} lists the extinction corrections, corrected line luminosities, FIR reference SFRs, and [O~{\sc ii}] SFRs used for the comparison in Section~\ref{sec:compareo2firsfr}.

\begin{table}[H]
\caption{Emission-line measurements for the SDSS quasar sample.}
\label{tab:115sdssspecresult}
\begin{adjustwidth}{-\extralength}{0cm}
\setlength{\tabcolsep}{4.43mm}
\begin{tabular}{rlccccc}
\toprule

\multirow{2}{*}{\textbf{{No.} 
}} & \multirow{2}{*}{\textbf{Name}} & \multirow{2}{*}{\boldmath$z$} &
\boldmath$\log L_{\rm [O\,II]}$ &
\boldmath$\log L_{\rm [O\,III]}$ &
{\mbox{\boldmath$\log L_{{\rm H}\alpha}$}} 
&
\boldmath$\log L_{{\rm H}\beta}$ \\
&  &  &
\textbf{(\boldmath${\rm erg~s^{-1}}$)} &
\textbf{(\boldmath${\rm erg~s^{-1}}$) }&
\textbf{(\boldmath${\rm erg~s^{-1}}$)} &
\textbf{(\boldmath${\rm erg~s^{-1}}$)} \\
\midrule

1&{J001115.69+011459.23} 
&0.580&41.60~$\pm$~0.10&41.28~$\pm$~0.36&--&41.51~$\pm$~0.08\\
2&J002421.84+002508.49&0.599&41.39~$\pm$~0.10&41.50~$\pm$~0.08&--&41.81~$\pm$~0.05\\
3&J005905.51+000651.67&0.719&42.46~$\pm$~0.02&43.11~$\pm$~0.02&--&42.50~$\pm$~0.02\\
4&J011430.24+000420.94&0.455&41.05~$\pm$~0.09&40.26~$\pm$~1.40&--&39.89~$\pm$~0.84\\
5&J011536.92{-}
000011.01&0.532&41.67~$\pm$~0.03&42.21~$\pm$~0.02&--&41.51~$\pm$~0.06\\
6&J013354.30+011034.00&0.704&41.78~$\pm$~0.05&41.86~$\pm$~0.11&--&41.48~$\pm$~0.35\\
7&J014003.99{-}
002541.77&0.770&42.08~$\pm$~0.03&41.89~$\pm$~0.27&--&41.46~$\pm$~0.25\\
8&J015243.15+002039.65&0.578&41.78~$\pm$~0.07&42.14~$\pm$~0.06&--&41.15~$\pm$~0.26\\
9&J015950.25+002340.82&0.163&41.35~$\pm$~0.02&42.26~$\pm$~0.01&42.54~$\pm$~0.01&41.59~$\pm$~0.02\\
10&J021859.87+002855.80&0.351&41.36~$\pm$~0.06&41.55~$\pm$~0.04&42.36~$\pm$~0.03&41.87~$\pm$~0.03\\
11&J022430.61{-}
000038.91&0.431&41.44~$\pm$~0.03&41.42~$\pm$~0.20&--&41.78~$\pm$~0.07\\
12&J090026.51+204158.70&0.706&41.96~$\pm$~0.09&42.16~$\pm$~1.65&--&--\\
13&J090158.88+002313.87&0.196&41.23~$\pm$~0.02&41.74~$\pm$~0.04&41.76~$\pm$~0.01&41.06~$\pm$~0.03\\
14&J090933.49+425346.51&0.670&42.65~$\pm$~0.01&42.25~$\pm$~0.13&--&41.62~$\pm$~0.22\\
15&J092159.40+450912.38&0.235&41.88~$\pm$~0.01&41.20~$\pm$~0.20&42.30~$\pm$~0.02&41.55~$\pm$~0.01\\
16&J092554.72+195405.13&0.192&40.76~$\pm$~0.23&42.09~$\pm$~0.01&42.26~$\pm$~0.03&41.84~$\pm$~0.03\\
17&J092635.12+072446.43&0.189&41.44~$\pm$~0.01&42.03~$\pm$~0.01&41.70~$\pm$~0.02&40.91~$\pm$~0.04\\

18&J094652.58+131953.83&0.133&42.12~$\pm$~0.01&42.44~$\pm$~0.04&42.83~$\pm$~0.01&42.01~$\pm$~0.03\\
19&J094745.14+072520.58&0.086&40.96~$\pm$~0.02&41.56~$\pm$~0.05&41.66~$\pm$~0.01&41.02~$\pm$~0.02\\

\bottomrule
\end{tabular}
\end{adjustwidth}
\end{table}

\begin{table}[H]\ContinuedFloat
\caption{{\em Cont.}}

\label{tab:115sdssspecresult}
\begin{adjustwidth}{-\extralength}{0cm}
\setlength{\tabcolsep}{4.48mm}
\begin{tabular}{rlccccc}
\toprule

\multirow{2}{*}{\textbf{No.}} & \multirow{2}{*}{\textbf{Name}} & \multirow{2}{*}{\boldmath$z$} &
\boldmath$\log L_{\rm [O\,II]}$ &
\boldmath$\log L_{\rm [O\,III]}$ &
\boldmath$\log L_{\rm H\alpha}$ &
\boldmath$\log L_{\rm H\beta}$ \\
&  &  &
\textbf{(\boldmath${\rm erg~s^{-1}}$)} &
\textbf{(\boldmath${\rm erg~s^{-1}}$) }&
\textbf{(\boldmath${\rm erg~s^{-1}}$)} &
\textbf{(\boldmath${\rm erg~s^{-1}}$)} \\
\midrule

20&J095017.06+215022.41&0.455&40.91~$\pm$~0.14&41.91~$\pm$~0.02&42.61~$\pm$~0.01&41.77~$\pm$~0.03\\

21&J095240.17+515249.91&0.554&41.24~$\pm$~0.13&42.25~$\pm$~0.02&--&40.70~$\pm$~0.58\\
22&J095819.87+022903.51&0.345&40.80~$\pm$~0.11&41.14~$\pm$~0.10&41.49~$\pm$~0.10&40.75~$\pm$~0.10\\
23&J100043.14+020637.25&0.360&41.20~$\pm$~0.04&41.38~$\pm$~0.15&41.90~$\pm$~0.03&40.70~$\pm$~0.08\\

24&J100232.13+023537.33&0.658&41.60~$\pm$~0.06&42.22~$\pm$~0.05&--&41.89~$\pm$~0.12\\
25&J100420.13+051300.48&0.160&41.10~$\pm$~0.05&41.81~$\pm$~0.01&42.80~$\pm$~0.01&42.05~$\pm$~0.01\\
26&J103651.94+575950.96&0.500&41.54~$\pm$~0.06&41.65~$\pm$~0.23&--&41.56~$\pm$~0.12\\

27&J104009.33+560343.28&0.393&41.44~$\pm$~0.05&42.07~$\pm$~0.03&42.56~$\pm$~0.04&41.88~$\pm$~0.05\\
28&J104505.39+561118.34&0.428&42.36~$\pm$~0.01&42.43~$\pm$~0.04&--&41.87~$\pm$~0.02\\

29&J104739.49+563507.19&0.303&40.98~$\pm$~0.08&41.18~$\pm$~0.18&41.58~$\pm$~0.11&40.96~$\pm$~0.08\\
30&J105106.12+591625.24&0.767&42.08~$\pm$~0.09&42.45~$\pm$~0.15&--&--\\
31&J105143.89+335926.71&0.167&41.33~$\pm$~0.04&42.06~$\pm$~0.10&42.45~$\pm$~0.01&41.23~$\pm$~0.04\\
32&J105151.44-005117.66&0.359&42.46~$\pm$~0.02&43.37~$\pm$~0.01&43.02~$\pm$~0.01&42.31~$\pm$~0.02\\
33&J105705.41+580437.46&0.140&40.62~$\pm$~0.05&41.35~$\pm$~0.02&41.48~$\pm$~0.02&40.60~$\pm$~0.13\\
34&J105959.93+574848.17&0.453&41.36~$\pm$~0.06&42.38~$\pm$~0.04&--&41.72~$\pm$~0.04\\
35&J110036.64+564134.89&0.834&41.58~$\pm$~0.12&--&--&--\\
36&J111706.40+441333.30&0.144&41.41~$\pm$~0.04&42.25~$\pm$~0.03&42.44~$\pm$~0.01&41.65~$\pm$~0.04\\
37&J111830.29+402554.02&0.154&41.13~$\pm$~0.06&41.63~$\pm$~0.09&42.84~$\pm$~0.01&42.17~$\pm$~0.02\\
38&J112019.62+130320.10&0.314&41.28~$\pm$~0.04&41.54~$\pm$~0.23&42.34~$\pm$~0.01&41.13~$\pm$~0.34\\
39&J112048.99+133821.91&0.513&41.52~$\pm$~0.05&41.83~$\pm$~0.07&--&41.38~$\pm$~0.06\\
40&J112759.26+360207.00&0.667&41.84~$\pm$~0.03&42.55~$\pm$~0.03&--&41.45~$\pm$~0.11\\
41&J114121.76+014803.58&0.382&41.62~$\pm$~0.04&42.09~$\pm$~0.03&42.46~$\pm$~0.02&41.73~$\pm$~0.03\\
42&J114933.88+222227.07&0.554&41.32~$\pm$~0.09&41.27~$\pm$~0.35&--&41.30~$\pm$~0.11\\
43&J115253.68+000131.90&0.823&41.91~$\pm$~0.07&42.35~$\pm$~0.07&--&41.51~$\pm$~0.50\\
44&J120226.76{-}
012915.28&0.150&41.25~$\pm$~0.02&41.36~$\pm$~0.01&42.02~$\pm$~0.01&41.10~$\pm$~0.03\\
45&J120312.14+015321.30&0.296&41.02~$\pm$~0.04&40.80~$\pm$~0.25&41.67~$\pm$~0.02&41.54~$\pm$~0.07\\
46&J120442.11+275411.80&0.165&41.69~$\pm$~0.01&42.32~$\pm$~0.01&42.22~$\pm$~0.01&41.59~$\pm$~0.01\\
47&J120734.63+150643.69&0.750&41.74~$\pm$~0.08&42.56~$\pm$~0.18&--&42.54~$\pm$~0.12\\
48&J121037.81+053805.88&0.436&41.09~$\pm$~0.11&41.75~$\pm$~0.08&--&41.44~$\pm$~0.12\\
49&J121728.37+065110.00&0.589&41.56~$\pm$~0.09&42.19~$\pm$~0.08&--&41.55~$\pm$~0.15\\
50&J121836.71+155908.43&0.766&42.01~$\pm$~0.05&42.33~$\pm$~0.10&--&42.45~$\pm$~0.07\\
51&J121906.57+160243.06&0.761&42.11~$\pm$~0.03&41.55~$\pm$~0.56&--&41.50~$\pm$~0.34\\
52&J121945.03+082117.95&0.228&41.18~$\pm$~0.02&40.86~$\pm$~0.10&42.05~$\pm$~0.01&41.07~$\pm$~0.03\\
53&J121946.54+145259.37&0.401&41.38~$\pm$~0.05&41.90~$\pm$~0.05&41.14~$\pm$~0.64&40.94~$\pm$~0.15\\
54&J122011.88+020342.21&0.240&41.95~$\pm$~0.02&42.79~$\pm$~0.01&42.89~$\pm$~0.01&42.04~$\pm$~0.02\\
55&J122026.72+062748.20&0.349&40.67~$\pm$~0.12&41.27~$\pm$~0.14&40.82~$\pm$~0.32&40.44~$\pm$~0.16\\
56&J122102.50+155447.04&0.229&41.17~$\pm$~0.04&41.81~$\pm$~0.01&42.08~$\pm$~0.03&41.30~$\pm$~0.05\\
57&J122102.95{-}
000733.74&0.366&41.94~$\pm$~0.01&41.64~$\pm$~0.03&42.69~$\pm$~0.02&41.96~$\pm$~0.01\\
58&J122106.50+114625.46&0.340&41.40~$\pm$~0.03&41.80~$\pm$~0.06&42.10~$\pm$~0.02&41.31~$\pm$~0.03\\
59&J122307.37-002124.73&0.804&42.09~$\pm$~0.03&42.68~$\pm$~0.02&--&41.40~$\pm$~0.34\\
60&J122312.17+095017.72&0.277&40.96~$\pm$~0.08&41.80~$\pm$~0.10&41.83~$\pm$~0.11&41.40~$\pm$~0.10\\
61&J122317.80+092306.94&0.682&41.94~$\pm$~0.10&41.86~$\pm$~0.15&--&42.48~$\pm$~0.04\\
62&J122404.62+045637.94&0.358&41.80~$\pm$~0.02&42.01~$\pm$~0.11&42.44~$\pm$~0.04&41.72~$\pm$~0.06\\
63&J122520.13+084450.76&0.535&41.94~$\pm$~0.06&42.31~$\pm$~0.05&--&42.02~$\pm$~0.07\\
64&J122526.21+141332.24&0.760&41.76~$\pm$~0.09&41.94~$\pm$~0.11&--&42.19~$\pm$~0.07\\
65&J122641.50+055906.81&0.290&41.11~$\pm$~0.06&41.74~$\pm$~0.16&41.78~$\pm$~0.02&40.99~$\pm$~0.07\\
66&J122822.10+114606.83&0.365&41.71~$\pm$~0.02&41.98~$\pm$~0.01&42.24~$\pm$~0.02&41.56~$\pm$~0.02\\
67&J122839.20+035749.29&0.608&41.35~$\pm$~0.07&41.57~$\pm$~0.36&--&40.94~$\pm$~0.47\\
68&J123436.54+123918.66&0.777&42.10~$\pm$~0.02&42.56~$\pm$~0.07&--&41.67~$\pm$~0.19\\
69&J123800.92+621336.09&0.440&41.15~$\pm$~0.09&41.61~$\pm$~0.03&--&41.65~$\pm$~0.06\\
70&J124511.26+335610.12&0.711&--&42.41~$\pm$~0.02&--&42.36~$\pm$~0.05\\
71&J125257.36+331555.81&0.367&41.50~$\pm$~0.03&41.63~$\pm$~0.06&41.84~$\pm$~0.11&41.09~$\pm$~0.09\\
72&J125317.57+310550.64&0.782&42.29~$\pm$~0.05&42.90~$\pm$~0.08&--&42.41~$\pm$~0.14\\

\bottomrule
\end{tabular}
\end{adjustwidth}
\end{table}

\begin{table}[H]\ContinuedFloat
\caption{{\em Cont.}}

\label{tab:115sdssspecresult}
\begin{adjustwidth}{-\extralength}{0cm}
\setlength{\tabcolsep}{4.48mm}
\begin{tabular}{rlccccc}
\toprule

\multirow{2}{*}{\textbf{No.}} & \multirow{2}{*}{\textbf{Name}} & \multirow{2}{*}{\boldmath$z$} &
\boldmath$\log L_{\rm [O\,II]}$ &
\boldmath$\log L_{\rm [O\,III]}$ &
\boldmath$\log L_{\rm H\alpha}$ &
\boldmath$\log L_{\rm H\beta}$ \\
&  &  &
\textbf{(\boldmath${\rm erg~s^{-1}}$)} &
\textbf{(\boldmath${\rm erg~s^{-1}}$) }&
\textbf{(\boldmath${\rm erg~s^{-1}}$)} &
\textbf{(\boldmath${\rm erg~s^{-1}}$)} \\
\midrule

73&J125553.05+272405.23&0.316&41.04~$\pm$~0.07&41.65~$\pm$~0.03&41.65~$\pm$~0.07&40.87~$\pm$~0.12\\
74&J125703.80+250457.56&0.821&41.65~$\pm$~0.09&--&--&--\\
75&J125711.97+274216.45&0.793&41.76~$\pm$~0.04&41.91~$\pm$~0.36&--&41.45~$\pm$~0.39\\
76&J125757.23+322929.29&0.806&41.39~$\pm$~0.22&42.44~$\pm$~0.03&--&41.72~$\pm$~0.22\\
77&J130622.96+225752.95&0.758&41.70~$\pm$~0.06&40.94~$\pm$~1.14&--&41.30~$\pm$~0.32\\
78&J130947.00+081948.23&0.154&41.71~$\pm$~0.03&42.57~$\pm$~0.03&42.66~$\pm$~0.01&41.84~$\pm$~0.02\\
79&J131247.97+250756.71&0.425&42.11~$\pm$~0.02&42.64~$\pm$~0.01&--&41.87~$\pm$~0.02\\
80&J131312.13+284730.01&0.259&40.83~$\pm$~0.07&40.85~$\pm$~0.12&--&40.31~$\pm$~0.15\\

81&J131531.66+265414.73&0.620&41.59~$\pm$~0.08&42.28~$\pm$~0.11&--&41.50~$\pm$~0.17\\
82&J132919.84+250626.45&0.781&41.39~$\pm$~0.13&41.58~$\pm$~0.25&--&--\\
83&J133005.71+254243.77&0.598&--&42.34~$\pm$~0.03&--&41.56~$\pm$~0.13\\
84&J133442.29+320939.60&0.649&41.24~$\pm$~0.20&41.51~$\pm$~0.22&--&42.28~$\pm$~0.06\\
85&J134356.74+253847.69&0.086&40.63~$\pm$~0.05&41.49~$\pm$~0.03&41.69~$\pm$~0.01&40.74~$\pm$~0.03\\
86&J135632.80+210352.35&0.301&41.58~$\pm$~0.06&42.57~$\pm$~0.01&42.43~$\pm$~0.03&41.79~$\pm$~0.03\\
87&J140621.89+222346.54&0.098&40.29~$\pm$~0.09&--&42.51~$\pm$~0.00&41.93~$\pm$~0.00\\
88&J140655.66+015712.88&0.427&41.40~$\pm$~0.04&41.93~$\pm$~0.02&--&41.88~$\pm$~0.05\\
89&J140700.40+282714.65&0.077&41.10~$\pm$~0.02&41.75~$\pm$~0.01&41.84~$\pm$~0.01&41.26~$\pm$~0.02\\
90&J141637.45+003352.28&0.434&41.83~$\pm$~0.02&42.55~$\pm$~0.04&--&41.74~$\pm$~0.03\\
91&J141644.62+190541.92&0.365&41.27~$\pm$~0.04&40.15~$\pm$~1.17&41.97~$\pm$~0.03&41.31~$\pm$~0.14\\
92&J141700.83+445606.39&0.113&40.45~$\pm$~0.07&41.25~$\pm$~0.04&42.25~$\pm$~0.01&--\\
93&J142052.44+525622.42&0.677&41.68~$\pm$~0.08&42.50~$\pm$~0.02&--&42.13~$\pm$~0.04\\
94&J142648.78+005323.24&0.220&41.50~$\pm$~0.01&41.92~$\pm$~0.16&42.20~$\pm$~0.01&41.21~$\pm$~0.04\\
95&J142710.94+013023.22&0.704&42.30~$\pm$~0.04&41.86~$\pm$~0.26&--&41.91~$\pm$~0.27\\
96&J142753.79+345248.35&0.514&41.26~$\pm$~0.12&41.75~$\pm$~0.04&--&42.18~$\pm$~0.03\\
97&J142918.15+592106.65&0.739&42.24~$\pm$~0.02&42.86~$\pm$~0.08&--&42.16~$\pm$~0.06\\
98&J142943.07+474726.23&0.221&41.53~$\pm$~0.02&42.51~$\pm$~0.01&42.91~$\pm$~0.00&41.95~$\pm$~0.01\\
99&J143624.81{-}
002905.36&0.325&40.93~$\pm$~0.10&41.98~$\pm$~0.01&42.70~$\pm$~0.01&41.73~$\pm$~0.03\\
100&J144231.82+014353.43&0.280&40.79~$\pm$~0.08&40.98~$\pm$~0.12&41.71~$\pm$~0.04&40.82~$\pm$~0.14\\
101&J145001.69+022006.75&0.520&41.75~$\pm$~0.04&--&--&41.81~$\pm$~0.08\\
102&J145108.76+270926.92&0.064&41.02~$\pm$~0.02&41.77~$\pm$~0.01&42.50~$\pm$~0.01&41.85~$\pm$~0.01\\
103&J145538.73+002238.06&0.434&40.83~$\pm$~0.16&41.61~$\pm$~0.05&--&41.19~$\pm$~0.09\\
104&J152114.26+222743.87&0.136&40.87~$\pm$~0.05&41.62~$\pm$~0.01&42.72~$\pm$~0.01&41.78~$\pm$~0.02\\
105&J154530.24+484608.98&0.399&41.68~$\pm$~0.05&41.89~$\pm$~0.03&43.48~$\pm$~0.00&42.75~$\pm$~0.01\\
106&J155444.58+082221.48&0.119&40.42~$\pm$~0.10&41.56~$\pm$~0.01&42.70~$\pm$~0.00&42.00~$\pm$~0.01\\
107&J161413.20+260416.21&0.131&42.24~$\pm$~0.01&43.00~$\pm$~0.02&42.83~$\pm$~0.00&42.01~$\pm$~0.01\\
108&J163352.34+402115.66&0.782&41.79~$\pm$~0.05&42.12~$\pm$~0.18&--&41.97~$\pm$~0.12\\
109&J163915.81+412833.70&0.690&41.66~$\pm$~0.13&42.26~$\pm$~0.04&--&41.53~$\pm$~0.15\\
110&J171033.22+584456.86&0.281&41.01~$\pm$~0.05&41.25~$\pm$~0.04&41.98~$\pm$~0.01&41.09~$\pm$~0.03\\
111&J171352.43+584201.25&0.521&42.03~$\pm$~0.03&42.74~$\pm$~0.02&--&41.63~$\pm$~0.07\\
112&J220759.39+001722.62&0.368&40.84~$\pm$~0.08&40.71~$\pm$~0.20&42.75~$\pm$~0.00&42.08~$\pm$~0.01\\
113&J223607.68+134355.32&0.326&41.49~$\pm$~0.05&42.58~$\pm$~0.01&43.32~$\pm$~0.00&42.62~$\pm$~0.07\\
114&J233741.32+001743.77&0.762&41.66~$\pm$~0.06&41.88~$\pm$~0.08&--&41.70~$\pm$~0.05\\
115&J235156.13{-}
010913.34&0.174&41.73~$\pm$~0.01&42.23~$\pm$~0.01&42.20~$\pm$~0.00&41.57~$\pm$~0.01\\
\bottomrule
\end{tabular}
\end{adjustwidth}
\footnotesize
\noindent\rev{{Notes:} 
Columns (4)--(7) list the logarithmic luminosities of the narrow emission-line components. A ``--'' entry indicates that the corresponding line is not covered or cannot be reliably measured. H$\alpha$ is generally outside the SDSS spectral coverage for sources at $z\gtrsim0.4$.}
\end{table}

\vspace{-12pt}

\begin{table}[H]
\caption{Emission-line measurements for PG quasars with SDSS Spectra.}
\label{tab:31pgsdssspecresult}
\begin{adjustwidth}{-\extralength}{0cm}

\setlength{\tabcolsep}{5.5mm}
\begin{tabular}{r l c c c c c}

\toprule
\multirow{2}{*}{\textbf{No.}} & \multirow{2}{*}{\textbf{Name}} & \multirow{2}{*}{\boldmath$z$} &
\boldmath$\log L_{\rm [O\,II]}$ &
\boldmath$\log L_{\rm [O\,III]}$ &
\boldmath$\log L_{\rm H\alpha}$ &
\boldmath$\log L_{\rm H\beta}$ \\
&  &  &
\textbf{(\boldmath${\rm erg~s^{-1}}$)} &
\textbf{(\boldmath${\rm erg~s^{-1}}$)} &
\textbf{(\boldmath${\rm erg~s^{-1}}$)} &
\textbf{(\boldmath${\rm erg~s^{-1}}$)} \\
\midrule

1&PG0157+001&0.164&41.41~$\pm$~0.02&42.26~$\pm$~0.01&42.55~$\pm$~0.00&41.60~$\pm$~0.02\\
2&PG0921+525&0.035&41.21~$\pm$~0.00&41.80~$\pm$~0.01&41.69~$\pm$~0.01&40.80~$\pm$~0.01\\

\bottomrule
\end{tabular}
\end{adjustwidth}
\end{table}

\vspace{-12pt}
\begin{table}[H]\ContinuedFloat
\caption{{\em Cont.}}

\label{tab:31pgsdssspecresult}
\begin{adjustwidth}{-\extralength}{0cm}

\setlength{\tabcolsep}{5.5mm}
\begin{tabular}{r l c c c c c}

\toprule
\multirow{2}{*}{\textbf{No.}} & \multirow{2}{*}{\textbf{Name}} & \multirow{2}{*}{\boldmath$z$} &
\boldmath$\log L_{\rm [O\,II]}$ &
\boldmath$\log L_{\rm [O\,III]}$ &
\boldmath$\log L_{\rm H\alpha}$ &
\boldmath$\log L_{\rm H\beta}$ \\
&  &  &
\textbf{(\boldmath${\rm erg~s^{-1}}$)} &
\textbf{(\boldmath${\rm erg~s^{-1}}$)} &
\textbf{(\boldmath${\rm erg~s^{-1}}$)} &
\textbf{(\boldmath${\rm erg~s^{-1}}$)} \\
\midrule

3&PG0923+201&0.190&41.18~$\pm$~0.11&41.32~$\pm$~0.05&40.99~$\pm$~0.35&41.26~$\pm$~0.11\\

4&PG0934+013&0.050&40.56~$\pm$~0.02&41.44~$\pm$~0.01&41.78~$\pm$~0.00&40.86~$\pm$~0.02\\
5&PG0947+396&0.206&41.32~$\pm$~0.07&42.08~$\pm$~0.01&42.36~$\pm$~0.01&41.33~$\pm$~0.04\\
6&PG1001+054&0.161&41.11~$\pm$~0.05&41.81~$\pm$~0.01&42.81~$\pm$~0.00&42.06~$\pm$~0.01\\

7&PG1004+130&0.240&41.53~$\pm$~0.12&42.43~$\pm$~0.01&42.50~$\pm$~0.01&41.95~$\pm$~0.03\\
8&PG1022+519&0.045&40.34~$\pm$~0.03&40.89~$\pm$~0.01&41.56~$\pm$~0.05&40.70~$\pm$~0.02\\

9&PG1048+342&0.167&41.11~$\pm$~0.04&41.96~$\pm$~0.01&42.55~$\pm$~0.00&41.33~$\pm$~0.02\\
10&PG1049{-}
005&0.357&42.45~$\pm$~0.02&43.36~$\pm$~0.01&41.98~$\pm$~0.11&42.31~$\pm$~0.02\\

11&PG1114+445&0.144&41.41~$\pm$~0.04&42.25~$\pm$~0.04&42.44~$\pm$~0.01&41.64~$\pm$~0.05\\

12&PG1115+407&0.154&41.17~$\pm$~0.06&41.80~$\pm$~0.01&42.84~$\pm$~0.01&42.06~$\pm$~0.02\\
13&PG1121+422&0.234&41.36~$\pm$~0.05&41.90~$\pm$~0.14&43.11~$\pm$~0.19&42.37~$\pm$~0.07\\
14&PG1151+117&0.176&41.25~$\pm$~0.06&41.91~$\pm$~0.08&42.02~$\pm$~0.04&41.35~$\pm$~0.17\\
15&PG1202+281&0.165&41.69~$\pm$~0.01&42.32~$\pm$~0.01&42.22~$\pm$~0.01&41.59~$\pm$~0.01\\
16&PG1229+204&0.064&40.81~$\pm$~0.02&41.19~$\pm$~0.02&41.81~$\pm$~0.01&41.18~$\pm$~0.01\\
17&PG1244+026&0.048&40.50~$\pm$~0.03&40.40~$\pm$~0.44&41.72~$\pm$~0.01&41.26~$\pm$~0.04\\
18&PG1259+593&0.472&--&--&--&--\\
19&PG1309+355&0.184&41.44~$\pm$~0.07&42.38~$\pm$~0.02&42.63~$\pm$~0.01&41.79~$\pm$~0.04\\
20&PG1341+258&0.087&40.73~$\pm$~0.04&41.41~$\pm$~0.05&41.69~$\pm$~0.01&40.66~$\pm$~0.05\\
21&PG1415+451&0.114&40.41~$\pm$~0.08&41.25~$\pm$~0.06&42.25~$\pm$~0.01&--\\
22&PG1425+267&0.366&42.16~$\pm$~0.02&42.87~$\pm$~0.02&42.67~$\pm$~0.01&42.01~$\pm$~0.02\\
23&PG1427+480&0.221&41.45~$\pm$~0.03&42.44~$\pm$~0.01&42.95~$\pm$~0.00&41.92~$\pm$~0.02\\
24&PG1444+407&0.267&40.98~$\pm$~0.33&42.17~$\pm$~0.01&43.08~$\pm$~0.01&42.42~$\pm$~0.02\\
25&PG1512+370&0.371&42.16~$\pm$~0.03&42.90~$\pm$~0.02&42.77~$\pm$~0.01&42.08~$\pm$~0.02\\
26&PG1534+580&0.030&40.63~$\pm$~0.01&41.60~$\pm$~0.01&41.36~$\pm$~0.01&40.61~$\pm$~0.01\\
27&PG1543+489&0.400&41.68~$\pm$~0.05&41.89~$\pm$~0.03&43.48~$\pm$~0.00&42.75~$\pm$~0.01\\
28&PG1552+085&0.119&40.43~$\pm$~0.12&41.56~$\pm$~0.01&42.70~$\pm$~0.01&42.01~$\pm$~0.01\\
29&PG1612+261&0.131&42.24~$\pm$~0.01&43.00~$\pm$~0.03&42.83~$\pm$~0.00&42.01~$\pm$~0.01\\
30&PG1704+608&0.371&42.53~$\pm$~0.02&43.36~$\pm$~0.01&45.80~$\pm$~0.22&42.47~$\pm$~0.01\\
31&PG2233+134&0.325&41.51~$\pm$~0.07&42.58~$\pm$~0.01&43.32~$\pm$~0.00&42.62~$\pm$~0.05\\
\bottomrule
\end{tabular}
\end{adjustwidth}
\footnotesize
\noindent{Notes:} 
The columns are the same as in Table~\ref{tab:115sdssspecresult}, but for the PG quasars with available SDSS spectra. Columns (4)--(7) give the logarithmic luminosities of the narrow emission-line components.
\end{table}

\vspace{-12pt}

\begin{table}[H]
\caption{Emission-line measurements for newly observed PG quasars.}
\label{tab:52pgobsspecresult}

\setlength{\tabcolsep}{3.6mm}
\begin{adjustwidth}{-\extralength}{0cm}

\begin{tabular}{r l c c c c c c}

\toprule
\multirow{2}{*}{\textbf{No. }}& \multirow{2}{*}{\textbf{Name}} & \multirow{2}{*}{\boldmath$z$} &
\boldmath$\log L_{\rm [O\,II]}$ &
\boldmath$\log L_{\rm [O\,III]}$ &
\boldmath$\log L_{\rm H\alpha}$ &
\boldmath$\log L_{\rm H\beta}$ &
\multirow{2}{*}{\textbf{Instrument}} \\
&  &  &
\textbf{(\boldmath${\rm erg~s^{-1}}$)} &
\textbf{(\boldmath${\rm erg~s^{-1}}$)} &
\textbf{(\boldmath${\rm erg~s^{-1}}$)} &
\textbf{(\boldmath${\rm erg~s^{-1}}$)} &
\\

\midrule

1&PG0003+158&0.450&42.33~$\pm$~0.07&43.60~$\pm$~0.01&--&43.00~$\pm$~0.02&LJT\\
2&PG0003+199&0.025&40.82~$\pm$~0.03&41.72~$\pm$~0.00&42.29~$\pm$~0.00&41.53~$\pm$~0.01&HCT\\
3&PG0007+106&0.089&41.83~$\pm$~0.01&42.57~$\pm$~0.00&42.11~$\pm$~0.01&41.53~$\pm$~0.01&HCT\\
4&PG0026+129&0.142&41.57~$\pm$~0.02&41.36~$\pm$~0.04&42.05~$\pm$~0.02&--&XLT\\
5&PG0043+039&0.384&41.83~$\pm$~0.08&41.87~$\pm$~0.14&--&--&XLT\\
6&PG0049+171&0.064&41.07~$\pm$~0.04&42.11~$\pm$~0.01&42.19~$\pm$~0.00&41.38~$\pm$~0.01&HCT\\
7&PG0050+124&0.061&40.28~$\pm$~0.20&41.24~$\pm$~0.05&42.61~$\pm$~0.00&41.71~$\pm$~0.01&LJT\\
8&PG0052+251&0.155&41.98~$\pm$~0.02&42.88~$\pm$~0.01&42.72~$\pm$~0.01&41.96~$\pm$~0.02&LJT\\
9&PG0804+761&0.100&41.68~$\pm$~0.05&41.95~$\pm$~0.04&43.00~$\pm$~0.01&42.29~$\pm$~0.01&LJT\\
10&PG0838+770&0.131&40.56~$\pm$~0.20&41.35~$\pm$~0.14&42.08~$\pm$~0.01&41.34~$\pm$~0.06&LJT\\
11&PG0844+349&0.064&40.87~$\pm$~0.15&41.69~$\pm$~0.01&42.28~$\pm$~0.00&41.47~$\pm$~0.02&LJT\\
12&PG0923+129&0.029&40.60~$\pm$~0.05&40.97~$\pm$~0.06&41.87~$\pm$~0.01&41.02~$\pm$~0.02&LJT\\
13&PG0953+414&0.239&--&42.19~$\pm$~0.03&43.32~$\pm$~0.11&--&XLT\\
14&PG1011{-}
040&0.058&40.09~$\pm$~0.12&41.16~$\pm$~0.01&41.30~$\pm$~0.01&40.84~$\pm$~0.02&HCT\\
15&PG1012+008&0.185&41.78~$\pm$~0.03&42.16~$\pm$~0.02&42.36~$\pm$~0.02&41.92~$\pm$~0.03&HCT\\
16&PG1048-090&0.344&42.14~$\pm$~0.16&43.33~$\pm$~0.01&--&42.59~$\pm$~0.03&XLT\\
17&PG1100+772&0.313&42.52~$\pm$~0.02&43.31~$\pm$~0.02&42.41~$\pm$~0.28&42.39~$\pm$~0.03&XLT\\
\bottomrule
\end{tabular}
\end{adjustwidth}
\end{table}

\vspace{-12pt}
\begin{table}[H]\ContinuedFloat
\caption{{\em Cont.}}

\label{tab:52pgobsspecresult}

\setlength{\tabcolsep}{3.6mm}
\begin{adjustwidth}{-\extralength}{0cm}

\begin{tabular}{r l c c c c c c}

\toprule
\multirow{2}{*}{\textbf{No. }}& \multirow{2}{*}{\textbf{Name}} & \multirow{2}{*}{\boldmath$z$} &
\boldmath$\log L_{\rm [O\,II]}$ &
\boldmath$\log L_{\rm [O\,III]}$ &
\boldmath$\log L_{\rm H\alpha}$ &
\boldmath$\log L_{\rm H\beta}$ &
\multirow{2}{*}{\textbf{Instrument}} \\
&  &  &
\textbf{(\boldmath${\rm erg~s^{-1}}$)} &
\textbf{(\boldmath${\rm erg~s^{-1}}$)} &
\textbf{(\boldmath${\rm erg~s^{-1}}$)} &
\textbf{(\boldmath${\rm erg~s^{-1}}$)} &
\\
\midrule

18&PG1103-006&0.425&42.08~$\pm$~0.13&42.65~$\pm$~0.16&--&42.30~$\pm$~0.16&XLT\\
19&PG1116+215&0.177&41.46~$\pm$~0.23&40.92~$\pm$~0.82&43.15~$\pm$~0.01&42.54~$\pm$~0.02&XLT\\
20&PG1119+120&0.049&--&--&42.10~$\pm$~0.01&41.08~$\pm$~0.05&LJT\\
21&PG1126-041&0.060&40.83~$\pm$~0.06&41.20~$\pm$~0.06&41.71~$\pm$~0.01&41.59~$\pm$~0.01&HCT\\
22&PG1149{-}
110&0.049&40.77~$\pm$~0.02&39.64~$\pm$~5.52&39.40~$\pm$~0.52&39.59~$\pm$~0.30&HCT\\
23&PG1211+143&0.085&41.35~$\pm$~0.03&--&43.01~$\pm$~0.00&--&LJT\\

24&PG1216+069&0.334&42.21~$\pm$~0.14&42.96~$\pm$~0.12&43.11~$\pm$~0.15&42.38~$\pm$~0.12&XLT\\
25&PG1226+023&0.158&42.40~$\pm$~0.09&41.38~$\pm$~0.45&43.61~$\pm$~0.01&43.11~$\pm$~0.02&XLT\\
26&PG1307+085&0.155&41.88~$\pm$~0.03&42.75~$\pm$~0.02&42.57~$\pm$~0.01&42.00~$\pm$~0.03&XLT\\

27&PG1310{-}
108&0.035&40.47~$\pm$~0.05&41.61~$\pm$~0.01&41.43~$\pm$~0.01&40.66~$\pm$~0.02&LJT\\
28&PG1322+659&0.168&41.44~$\pm$~0.07&42.10~$\pm$~0.01&42.49~$\pm$~0.01&40.84~$\pm$~0.58&XLT\\
29&PG1351+640&0.087&41.69~$\pm$~0.04&42.59~$\pm$~0.01&42.72~$\pm$~0.00&42.10~$\pm$~0.01&HCT\\
30&PG1352+183&0.158&40.84~$\pm$~0.14&40.94~$\pm$~0.09&--&--&XLT\\
31&PG1354+213&0.300&41.88~$\pm$~0.07&--&42.62~$\pm$~0.02&42.08~$\pm$~0.07&XLT\\
32&PG1402+261&0.164&41.65~$\pm$~0.07&42.06~$\pm$~0.02&42.97~$\pm$~0.00&42.16~$\pm$~0.03&XLT\\
33&PG1411+442&0.089&41.18~$\pm$~0.24&42.28~$\pm$~0.01&42.70~$\pm$~0.01&42.07~$\pm$~0.03&HCT\\
34&PG1416{-}
129&0.129&41.60~$\pm$~0.01&41.93~$\pm$~0.06&41.95~$\pm$~0.01&41.34~$\pm$~0.03&LJT\\
35&PG1426+015&0.086&41.14~$\pm$~0.18&41.37~$\pm$~0.21&41.82~$\pm$~0.04&40.65~$\pm$~0.45&LJT\\
36&PG1440+356&0.077&41.08~$\pm$~0.10&41.26~$\pm$~0.27&42.61~$\pm$~0.01&41.67~$\pm$~0.03&LJT\\
37&PG1448+273&0.065&41.35~$\pm$~0.04&41.66~$\pm$~0.19&41.65~$\pm$~0.05&41.55~$\pm$~0.04&LJT\\
38&PG1501+106&0.036&41.09~$\pm$~0.02&42.00~$\pm$~0.00&41.82~$\pm$~0.00&41.22~$\pm$~0.01&LJT\\
39&PG1519+226&0.137&40.66~$\pm$~0.13&41.40~$\pm$~0.02&42.58~$\pm$~0.00&41.65~$\pm$~0.04&LJT\\
40&PG1535+547&0.038&--&40.86~$\pm$~0.01&42.00~$\pm$~0.00&41.23~$\pm$~0.01&LJT\\
41&PG1545+210&0.266&42.23~$\pm$~0.04&43.20~$\pm$~0.02&43.15~$\pm$~0.01&42.52~$\pm$~0.03&LJT\\
42&PG1613+658&0.129&41.71~$\pm$~0.03&42.24~$\pm$~0.02&41.83~$\pm$~0.03&40.57~$\pm$~0.37&LJT\\
43&PG1617+175&0.114&41.00~$\pm$~0.07&--&42.24~$\pm$~0.01&41.09~$\pm$~0.09&DBSP\\
44&PG1626+554&0.133&40.63~$\pm$~0.15&--&42.37~$\pm$~0.01&41.58~$\pm$~0.07&LJT\\
45&PG1700+518&0.282&--&--&43.65~$\pm$~0.00&--&LJT\\
46&PG2112+059&0.466&42.23~$\pm$~0.05&42.63~$\pm$~0.02&--&41.41~$\pm$~0.39&XLT\\
47&PG2130+099&0.061&41.02~$\pm$~0.03&41.91~$\pm$~0.01&42.75~$\pm$~0.00&41.94~$\pm$~0.01&LJT\\
48&PG2209+184&0.070&40.98~$\pm$~0.09&40.57~$\pm$~0.63&--&--&HCT\\
49&PG2214+139&0.067&41.15~$\pm$~0.10&41.28~$\pm$~0.02&41.89~$\pm$~0.01&--&HCT\\
50&PG2251+113&0.323&42.42~$\pm$~0.03&43.32~$\pm$~0.01&43.21~$\pm$~0.01&42.65~$\pm$~0.02&XLT\\
51&PG2304+042&0.042&40.64~$\pm$~0.05&41.59~$\pm$~0.01&41.29~$\pm$~0.03&40.51~$\pm$~0.04&HCT\\
52&PG2308+098&0.432&42.00~$\pm$~0.10&43.08~$\pm$~0.04&--&42.57~$\pm$~0.03&XLT\\
\bottomrule
\end{tabular}
\end{adjustwidth}
\footnotesize
\noindent\rev{{Notes:} 
Columns (4)--(7) list the logarithmic luminosities of the narrow emission-line components. The last column gives the instrument used for the optical spectroscopic observation.}
\end{table}

\vspace{-12pt}
\begin{table}[H]
\footnotesize
\caption{Extinction corrections and [O~{\sc ii}] star formation rates.}

\label{tab:o2sfrresult}
\setlength{\tabcolsep}{2.27mm}
\begin{adjustwidth}{-\extralength}{0cm}

\begin{tabular}{m{1.5cm}<{\raggedleft} l c c c c c c c}

\toprule
\multirow{2}{*}{\textbf{Sample {ID} 
}} & \multirow{2}{*}{\textbf{Name}} & \multirow{2}{*}{\boldmath$z$} &
\boldmath$\log {\rm SFR}_{\rm FIR}$ &
\multirow{2}{*}{\boldmath$E(B-V)_{\rm gas}$} &
\boldmath$\log L_{\rm [O\,II]}^{\rm corr}$ &
\boldmath$\log L_{\rm [O\,III]}^{\rm corr}$ &
\boldmath$\log {\rm SFR}_{\rm [O\,II]}$ &
\boldmath$\log {\rm SFR}_{\rm [O\,II]}^{\rm noAGNcorr}$ \\
&  &  &
\textbf{(\boldmath$M_\odot~{\rm yr}^{-1}$)} &
&
\textbf{(\boldmath${\rm erg~s^{-1}}$) }&
\textbf{(\boldmath${\rm erg~s^{-1}}$)} &
\textbf{(\boldmath$M_\odot~{\rm yr}^{-1}$)} &
\textbf{(\boldmath$M_\odot~{\rm yr}^{-1}$)} \\
\midrule

1&J001115.69+011459.23&0.580&2.34~$\pm$~0.15 &0.86~$\pm$~0.04&43.20~$\pm$~0.17&42.47~$\pm$~0.42&2.10~$\pm$~0.18&2.11~$\pm$~0.17\\
2&J002421.84+002508.49&0.599&2.12~$\pm$~0.15 &0.59~$\pm$~0.05&42.50~$\pm$~0.20&42.32~$\pm$~0.15&1.30~$\pm$~0.21&1.34~$\pm$~0.20\\
3&J005905.51+000651.67&0.719&2.00~$\pm$~0.38 &0.72~$\pm$~0.11&43.80~$\pm$~0.23&44.10~$\pm$~0.17&2.58~$\pm$~0.24&2.69~$\pm$~0.23\\
4&J011430.24+000420.94&0.455&1.43~$\pm$~0.14 &0.89~$\pm$~0.02&42.72~$\pm$~0.12&41.50~$\pm$~1.42&1.60~$\pm$~0.13&1.61~$\pm$~0.12\\
5&J011536.92-000011.01&0.532&1.72~$\pm$~0.14 &0.90~$\pm$~0.00&43.35~$\pm$~0.04&43.46~$\pm$~0.02&2.17~$\pm$~0.04&2.24~$\pm$~0.04\\
6&J013354.30+011034.00&0.704&1.92~$\pm$~0.24 &0.67~$\pm$~0.12&43.04~$\pm$~0.28&42.79~$\pm$~0.28&1.90~$\pm$~0.29&1.93~$\pm$~0.28\\
7&J014003.99-002541.77&0.770&2.40~$\pm$~0.19 &0.53~$\pm$~0.09&43.07~$\pm$~0.20&42.62~$\pm$~0.39&1.95~$\pm$~0.21&1.97~$\pm$~0.20\\
8&J015243.15+002039.65&0.578&1.48~$\pm$~0.26 &0.89~$\pm$~0.02&43.45~$\pm$~0.10&43.38~$\pm$~0.09&2.29~$\pm$~0.12&2.33~$\pm$~0.10\\
9&J015950.25+002340.82&0.163&2.40~$\pm$~0.12 &0.91~$\pm$~0.04&43.04~$\pm$~0.09&43.51~$\pm$~0.06&1.74~$\pm$~0.11&1.92~$\pm$~0.09\\
10&J021859.87+002855.80&0.351&1.96~$\pm$~0.18 &0.37~$\pm$~0.03&42.06~$\pm$~0.11&42.07~$\pm$~0.08&0.83~$\pm$~0.12&0.88~$\pm$~0.11\\
11&J022430.61-000038.91&0.431&1.41~$\pm$~0.13 &0.64~$\pm$~0.11&42.63~$\pm$~0.23&42.31~$\pm$~0.35&1.34~$\pm$~0.24&1.37~$\pm$~0.23\\
12&J090026.51+204158.70&0.706&2.35~$\pm$~0.06 &0.20~$\pm$~0.01&42.34~$\pm$~0.10&42.45~$\pm$~1.67&1.16~$\pm$~0.39&1.23~$\pm$~0.10\\

13&J090158.88+002313.87&0.196&1.52~$\pm$~0.09 &0.42~$\pm$~0.06&42.01~$\pm$~0.13&42.32~$\pm$~0.13&0.65~$\pm$~0.15&0.76~$\pm$~0.13\\

\bottomrule
\end{tabular}
\end{adjustwidth}
\end{table}

\begin{table}[H]\ContinuedFloat
\caption{{\em Cont.}}
\label{tab:o2sfrresult}
\setlength{\tabcolsep}{2.27mm}
\footnotesize
\begin{adjustwidth}{-\extralength}{0cm}

\begin{tabular}{m{1.5cm}<{\raggedleft} l c c c c c c c}

\toprule
\multirow{2}{*}{\textbf{Sample {ID} 
}} & \multirow{2}{*}{\textbf{Name}} & \multirow{2}{*}{\boldmath$z$} &
\boldmath$\log {\rm SFR}_{\rm FIR}$ &
\multirow{2}{*}{\boldmath$E(B-V)_{\rm gas}$} &
\boldmath$\log L_{\rm [O\,II]}^{\rm corr}$ &
\boldmath$\log L_{\rm [O\,III]}^{\rm corr}$ &
\boldmath$\log {\rm SFR}_{\rm [O\,II]}$ &
\boldmath$\log {\rm SFR}_{\rm [O\,II]}^{\rm noAGNcorr}$ \\
&  &  &
\textbf{(\boldmath$M_\odot~{\rm yr}^{-1}$)} &
&
\textbf{(\boldmath${\rm erg~s^{-1}}$) }&
\textbf{(\boldmath${\rm erg~s^{-1}}$)} &
\textbf{(\boldmath$M_\odot~{\rm yr}^{-1}$)} &
\textbf{(\boldmath$M_\odot~{\rm yr}^{-1}$)} \\
\midrule

14&J090933.49+425346.51&0.670&2.46~$\pm$~0.14 &0.75~$\pm$~0.02&44.05~$\pm$~0.05&43.28~$\pm$~0.16&2.93~$\pm$~0.05&2.94~$\pm$~0.05\\
15&J092159.40+450912.38&0.235&1.88~$\pm$~0.07 &0.59~$\pm$~0.02&42.98~$\pm$~0.05&42.02~$\pm$~0.23&1.73~$\pm$~0.05&1.73~$\pm$~0.05\\
17&J092635.12+072446.43&0.189&1.25~$\pm$~0.11 &0.60~$\pm$~0.09&42.55~$\pm$~0.18&42.86~$\pm$~0.13&1.33~$\pm$~0.18&1.44~$\pm$~0.18\\
18&J094652.58+131953.83&0.133&1.20~$\pm$~0.19 &0.64~$\pm$~0.06&43.31~$\pm$~0.13&43.32~$\pm$~0.13&2.11~$\pm$~0.13&2.16~$\pm$~0.13\\
19&J094745.14+072520.58&0.086&0.04~$\pm$~0.12 &0.30~$\pm$~0.05&41.51~$\pm$~0.10&41.97~$\pm$~0.11&0.25~$\pm$~0.13&0.42~$\pm$~0.10\\
20&J095017.06+215022.41&0.455&1.67~$\pm$~0.07 &0.68~$\pm$~0.07&42.19~$\pm$~0.26&42.85~$\pm$~0.11&0.72~$\pm$~0.43&1.03~$\pm$~0.26\\
21&J095240.17+515249.91&0.554&1.63~$\pm$~0.19 &0.32~$\pm$~0.17&41.84~$\pm$~0.45&42.69~$\pm$~0.26&0.11~$\pm$~0.98&0.75~$\pm$~0.45\\
23&J100043.14+020637.25&0.360&1.14~$\pm$~0.11 &0.43~$\pm$~0.05&42.01~$\pm$~0.13&41.98~$\pm$~0.22&0.86~$\pm$~0.15&0.91~$\pm$~0.13\\
24&J100232.13+023537.33&0.658&1.75~$\pm$~0.11 &0.45~$\pm$~0.18&42.45~$\pm$~0.40&42.85~$\pm$~0.30&1.21~$\pm$~0.44&1.35~$\pm$~0.40\\
25&J100420.13+051300.48&0.160&0.93~$\pm$~0.22 &0.52~$\pm$~0.03&42.07~$\pm$~0.10&42.53~$\pm$~0.05&0.76~$\pm$~0.13&0.92~$\pm$~0.10\\

26&J103651.94+575950.96&0.500&1.65~$\pm$~0.13 &0.50~$\pm$~0.11&42.47~$\pm$~0.26&42.34~$\pm$~0.39&1.31~$\pm$~0.29&1.35~$\pm$~0.26\\
28&J104505.39+561118.34&0.428&1.88~$\pm$~0.10 &0.75~$\pm$~0.02&43.76~$\pm$~0.05&43.46~$\pm$~0.07&2.62~$\pm$~0.05&2.65~$\pm$~0.05\\
29&J104739.49+563507.19&0.303&1.71~$\pm$~0.15 &0.59~$\pm$~0.09&42.08~$\pm$~0.25&41.99~$\pm$~0.30&0.88~$\pm$~0.27&0.92~$\pm$~0.25\\
30&J105106.12+591625.24&0.767&2.03~$\pm$~0.16 &0.20~$\pm$~0.10&42.46~$\pm$~0.27&42.73~$\pm$~0.28&1.26~$\pm$~0.34&1.36~$\pm$~0.27\\
31&J105143.89+335926.71&0.167&0.86~$\pm$~0.11 &1.45~$\pm$~0.07&44.03~$\pm$~0.17&44.07~$\pm$~0.20&2.86~$\pm$~0.19&2.92~$\pm$~0.17\\
32&J105151.44-005117.66&0.359&2.34~$\pm$~0.12 &0.44~$\pm$~0.05&43.28~$\pm$~0.12&43.98~$\pm$~0.09&1.84~$\pm$~0.16&2.19~$\pm$~0.12\\
33&J105705.41+580437.46&0.140&1.17~$\pm$~0.12 &0.80~$\pm$~0.05&42.12~$\pm$~0.14&42.46~$\pm$~0.09&0.89~$\pm$~0.16&1.01~$\pm$~0.14\\
34&J105959.93+574848.17&0.453&1.58~$\pm$~0.18 &0.44~$\pm$~0.09&42.19~$\pm$~0.23&42.99~$\pm$~0.16&0.45~$\pm$~0.46&0.96~$\pm$~0.23\\
36&J111706.40+441333.30&0.144&0.81~$\pm$~0.13 &0.60~$\pm$~0.09&42.53~$\pm$~0.20&43.09~$\pm$~0.15&1.22~$\pm$~0.25&1.44~$\pm$~0.20\\
37&J111830.29+402554.02&0.154&1.41~$\pm$~0.05 &0.36~$\pm$~0.04&41.80~$\pm$~0.12&42.14~$\pm$~0.14&0.48~$\pm$~0.17&0.60~$\pm$~0.12\\
38&J112019.62+130320.10&0.314&1.46~$\pm$~0.12 &0.52~$\pm$~0.08&42.24~$\pm$~0.18&42.26~$\pm$~0.34&1.03~$\pm$~0.22&1.08~$\pm$~0.18\\
39&J112048.99+133821.91&0.513&1.60~$\pm$~0.05 &0.34~$\pm$~0.12&42.15~$\pm$~0.28&42.30~$\pm$~0.24&0.85~$\pm$~0.30&0.93~$\pm$~0.28\\
40&J112759.26+360207.00&0.667&2.19~$\pm$~0.26 &0.63~$\pm$~0.14&43.03~$\pm$~0.29&43.43~$\pm$~0.22&1.79~$\pm$~0.31&1.92~$\pm$~0.29\\
41&J114121.76+014803.58&0.382&1.82~$\pm$~0.14 &0.48~$\pm$~0.07&42.50~$\pm$~0.17&42.75~$\pm$~0.13&1.25~$\pm$~0.19&1.35~$\pm$~0.17\\
42&J114933.88+222227.07&0.554&1.66~$\pm$~0.07 &0.21~$\pm$~0.03&41.71~$\pm$~0.15&41.56~$\pm$~0.39&0.52~$\pm$~0.19&0.55~$\pm$~0.15\\
43&J115253.68+000131.90&0.823&2.23~$\pm$~0.25 &0.69~$\pm$~0.12&43.20~$\pm$~0.30&43.31~$\pm$~0.25&2.03~$\pm$~0.32&2.09~$\pm$~0.30\\
44&J120226.76-012915.28&0.150&2.24~$\pm$~0.10 &0.85~$\pm$~0.07&42.84~$\pm$~0.15&42.55~$\pm$~0.11&1.65~$\pm$~0.15&1.67~$\pm$~0.15\\
45&J120312.14+015321.30&0.296&1.45~$\pm$~0.21 &0.81~$\pm$~0.08&42.52~$\pm$~0.19&41.92~$\pm$~0.36&1.41~$\pm$~0.20&1.43~$\pm$~0.19\\
46&J120442.11+275411.80&0.165&1.39~$\pm$~0.14 &0.27~$\pm$~0.03&42.19~$\pm$~0.06&42.70~$\pm$~0.05&0.80~$\pm$~0.07&0.99~$\pm$~0.06\\
47&J120734.63+150643.69&0.750&2.07~$\pm$~0.24 &0.35~$\pm$~0.09&42.41~$\pm$~0.25&43.05~$\pm$~0.31&1.02~$\pm$~0.49&1.30~$\pm$~0.25\\
48&J121037.81+053805.88&0.436&1.63~$\pm$~0.09 &0.43~$\pm$~0.08&41.90~$\pm$~0.25&42.36~$\pm$~0.18&0.53~$\pm$~0.34&0.69~$\pm$~0.25\\
49&J121728.37+065110.00&0.589&1.94~$\pm$~0.42 &0.13~$\pm$~0.03&41.79~$\pm$~0.15&42.36~$\pm$~0.13&0.47~$\pm$~0.27&0.69~$\pm$~0.15\\
50&J121836.71+155908.43&0.766&2.15~$\pm$~0.11 &0.37~$\pm$~0.05&42.70~$\pm$~0.15&42.85~$\pm$~0.17&1.48~$\pm$~0.18&1.55~$\pm$~0.15\\
51&J121906.57+160243.06&0.761&2.28~$\pm$~0.21 &0.63~$\pm$~0.10&43.28~$\pm$~0.22&42.41~$\pm$~0.71&2.16~$\pm$~0.23&2.17~$\pm$~0.22\\
52&J121945.03+082117.95&0.228&1.78~$\pm$~0.13 &0.96~$\pm$~0.07&42.98~$\pm$~0.15&42.20~$\pm$~0.19&1.84~$\pm$~0.15&1.85~$\pm$~0.15\\
53&J121946.54+145259.37&0.401&1.17~$\pm$~0.31 &0.05~$\pm$~0.01&41.47~$\pm$~0.06&41.97~$\pm$~0.05&0.00~$\pm$~0.11&0.19~$\pm$~0.06\\
54&J122011.88+020342.21&0.240&1.43~$\pm$~0.32 &0.71~$\pm$~0.04&43.28~$\pm$~0.09&43.78~$\pm$~0.06&1.99~$\pm$~0.11&2.17~$\pm$~0.09\\
55&J122026.72+062748.20&0.349&1.64~$\pm$~0.09 &0.83~$\pm$~0.05&42.22~$\pm$~0.22&42.43~$\pm$~0.21&1.05~$\pm$~0.27&1.13~$\pm$~0.22\\
56&J122102.50+155447.04&0.229&1.18~$\pm$~0.21 &0.57~$\pm$~0.12&42.23~$\pm$~0.25&42.60~$\pm$~0.17&0.90~$\pm$~0.27&1.03~$\pm$~0.25\\
57&J122102.95-000733.74&0.366&2.21~$\pm$~0.10 &0.46~$\pm$~0.05&42.81~$\pm$~0.10&42.28~$\pm$~0.09&1.70~$\pm$~0.10&1.71~$\pm$~0.10\\
58&J122106.50+114625.46&0.340&1.52~$\pm$~0.07 &0.59~$\pm$~0.08&42.50~$\pm$~0.18&42.62~$\pm$~0.16&1.32~$\pm$~0.19&1.39~$\pm$~0.18\\
59&J122307.37-002124.73&0.804&2.23~$\pm$~0.23 &0.88~$\pm$~0.03&43.74~$\pm$~0.09&43.91~$\pm$~0.06&2.55~$\pm$~0.10&2.63~$\pm$~0.09\\
60&J122312.17+095017.72&0.277&1.55~$\pm$~0.28 &0.39~$\pm$~0.10&41.69~$\pm$~0.27&42.34~$\pm$~0.24&0.22~$\pm$~0.45&0.52~$\pm$~0.27\\
61&J122317.80+092306.94&0.682&2.76~$\pm$~0.10 &0.25~$\pm$~0.02&42.40~$\pm$~0.13&42.20~$\pm$~0.18&1.27~$\pm$~0.15&1.30~$\pm$~0.13\\
62&J122404.62+045637.94&0.358&1.76~$\pm$~0.24 &0.45~$\pm$~0.14&42.64~$\pm$~0.28&42.64~$\pm$~0.30&1.42~$\pm$~0.30&1.47~$\pm$~0.28\\
63&J122520.13+084450.76&0.535&2.06~$\pm$~0.04 &0.68~$\pm$~0.14&43.20~$\pm$~0.32&43.25~$\pm$~0.25&2.03~$\pm$~0.34&2.09~$\pm$~0.32\\
64&J122526.21+141332.24&0.760&2.35~$\pm$~0.12 &0.31~$\pm$~0.10&42.34~$\pm$~0.28&42.37~$\pm$~0.26&1.18~$\pm$~0.30&1.24~$\pm$~0.28\\
65&J122641.50+055906.81&0.290&1.31~$\pm$~0.18 &0.90~$\pm$~0.01&42.78~$\pm$~0.08&42.98~$\pm$~0.17&1.59~$\pm$~0.13&1.67~$\pm$~0.08\\
66&J122822.10+114606.83&0.365&1.86~$\pm$~0.07 &0.39~$\pm$~0.06&42.43~$\pm$~0.13&42.52~$\pm$~0.09&1.26~$\pm$~0.13&1.32~$\pm$~0.13\\
67&J122839.20+035749.29&0.608&2.20~$\pm$~0.10 &0.27~$\pm$~0.02&41.85~$\pm$~0.12&41.94~$\pm$~0.39&0.68~$\pm$~0.19&0.75~$\pm$~0.12\\
68&J123436.54+123918.66&0.777&2.44~$\pm$~0.16 &0.44~$\pm$~0.06&42.93~$\pm$~0.13&43.17~$\pm$~0.15&1.73~$\pm$~0.15&1.82~$\pm$~0.13\\
69&J123800.92+621336.09&0.440&1.84~$\pm$~0.16 &0.47~$\pm$~0.05&42.03~$\pm$~0.18&42.27~$\pm$~0.10&0.80~$\pm$~0.20&0.89~$\pm$~0.18\\
71&J125257.36+331555.81&0.367&1.32~$\pm$~0.16 &0.89~$\pm$~0.02&43.15~$\pm$~0.07&42.86~$\pm$~0.09&2.02~$\pm$~0.08&2.04~$\pm$~0.07\\
72&J125317.57+310550.64&0.782&2.39~$\pm$~0.21 &0.08~$\pm$~0.08&42.45~$\pm$~0.19&43.02~$\pm$~0.19&1.11~$\pm$~0.28&1.34~$\pm$~0.19\\
73&J125553.05+272405.23&0.316&1.03~$\pm$~0.14 &0.86~$\pm$~0.04&42.64~$\pm$~0.14&42.83~$\pm$~0.09&1.44~$\pm$~0.16&1.53~$\pm$~0.14\\
75&J125711.97+274216.45&0.793&2.09~$\pm$~0.14 &0.17~$\pm$~0.03&42.08~$\pm$~0.10&42.15~$\pm$~0.40&0.91~$\pm$~0.16&0.97~$\pm$~0.10\\
76&J125757.23+322929.29&0.806&2.60~$\pm$~0.26 &0.74~$\pm$~0.05&42.76~$\pm$~0.32&43.46~$\pm$~0.10&1.31~$\pm$~0.61&1.65~$\pm$~0.32\\
77&J130622.96+225752.95&0.758&2.06~$\pm$~0.31 &0.47~$\pm$~0.09&42.57~$\pm$~0.24&41.59~$\pm$~1.27&1.47~$\pm$~0.25&1.48~$\pm$~0.24\\
78&J130947.00+081948.23&0.154&1.00~$\pm$~0.22 &0.64~$\pm$~0.04&42.91~$\pm$~0.11&43.47~$\pm$~0.09&1.61~$\pm$~0.14&1.82~$\pm$~0.11\\
79&J131247.97+250756.71&0.425&2.21~$\pm$~0.21 &0.78~$\pm$~0.09&43.56~$\pm$~0.19&43.72~$\pm$~0.14&2.38~$\pm$~0.19&2.45~$\pm$~0.19\\
80&J131312.13+284730.01&0.259&0.93~$\pm$~0.17 &0.89~$\pm$~0.03&42.48~$\pm$~0.11&42.08~$\pm$~0.16&1.35~$\pm$~0.12&1.37~$\pm$~0.11\\
81&J131531.66+265414.73&0.620&1.92~$\pm$~0.21 &0.61~$\pm$~0.12&42.72~$\pm$~0.30&43.12~$\pm$~0.27&1.42~$\pm$~0.37&1.56~$\pm$~0.30\\
82&J132919.84+250626.45&0.781&2.22~$\pm$~0.20 &0.40~$\pm$~0.10&42.13~$\pm$~0.31&42.13~$\pm$~0.39&0.96~$\pm$~0.35&1.01~$\pm$~0.31\\
84&J133442.29+320939.60&0.649&2.39~$\pm$~0.30 &0.20~$\pm$~0.01&41.61~$\pm$~0.23&41.79~$\pm$~0.23&0.44~$\pm$~0.31&0.52~$\pm$~0.23\\
85&J134356.74+253847.69&0.086&0.45~$\pm$~0.08 &0.91~$\pm$~0.07&42.33~$\pm$~0.18&42.75~$\pm$~0.13&1.02~$\pm$~0.21&1.17~$\pm$~0.18\\
86&J135632.80+210352.35&0.301&1.28~$\pm$~0.21 &0.29~$\pm$~0.09&42.12~$\pm$~0.23&42.97~$\pm$~0.13&0.19~$\pm$~0.46&0.82~$\pm$~0.23\\
88&J140655.66+015712.88&0.427&2.10~$\pm$~0.19 &0.36~$\pm$~0.08&42.08~$\pm$~0.20&42.43~$\pm$~0.14&0.81~$\pm$~0.22&0.93~$\pm$~0.20\\
90&J141637.45+003352.28&0.434&2.32~$\pm$~0.19 &0.69~$\pm$~0.04&43.11~$\pm$~0.10&43.51~$\pm$~0.10&1.86~$\pm$~0.12&1.99~$\pm$~0.10\\

\bottomrule
\end{tabular}
\end{adjustwidth}
\end{table}

\begin{table}[H]\ContinuedFloat
\caption{{\em Cont.}}
\label{tab:o2sfrresult}
\setlength{\tabcolsep}{2.13mm}
\footnotesize
\begin{adjustwidth}{-\extralength}{0cm}

\begin{tabular}{m{1.5cm}<{\raggedleft} l c c c c c c c}

\toprule
\multirow{2}{*}{\textbf{Sample {ID} 
}} & \multirow{2}{*}{\textbf{Name}} & \multirow{2}{*}{\boldmath$z$} &
\boldmath$\log {\rm SFR}_{\rm FIR}$ &
\multirow{2}{*}{\boldmath$E(B-V)_{\rm gas}$} &
\boldmath$\log L_{\rm [O\,II]}^{\rm corr}$ &
\boldmath$\log L_{\rm [O\,III]}^{\rm corr}$ &
\boldmath$\log {\rm SFR}_{\rm [O\,II]}$ &
\boldmath$\log {\rm SFR}_{\rm [O\,II]}^{\rm noAGNcorr}$ \\
&  &  &
\textbf{(\boldmath$M_\odot~{\rm yr}^{-1}$)} &
&
\textbf{(\boldmath${\rm erg~s^{-1}}$) }&
\textbf{(\boldmath${\rm erg~s^{-1}}$)} &
\textbf{(\boldmath$M_\odot~{\rm yr}^{-1}$)} &
\textbf{(\boldmath$M_\odot~{\rm yr}^{-1}$)} \\
\midrule

91&J141644.62+190541.92&0.365&1.67~$\pm$~0.11 &0.37~$\pm$~0.06&41.96~$\pm$~0.15&40.67~$\pm$~1.25&0.70~$\pm$~0.15&0.70~$\pm$~0.15\\
92&J141700.83+445606.39&0.113&0.73~$\pm$~0.27 &0.68~$\pm$~0.07&41.72~$\pm$~0.21&42.20~$\pm$~0.14&0.43~$\pm$~0.26&0.61~$\pm$~0.21\\
93&J142052.44+525622.42&0.677&2.17~$\pm$~0.29 &0.21~$\pm$~0.10&42.07~$\pm$~0.26&42.79~$\pm$~0.15&0.60~$\pm$~0.40&0.97~$\pm$~0.26\\
94&J142648.78+005323.24&0.220&1.40~$\pm$~0.11 &0.64~$\pm$~0.08&42.69~$\pm$~0.16&42.80~$\pm$~0.27&1.51~$\pm$~0.19&1.58~$\pm$~0.16\\
95&J142710.94+013023.22&0.704&2.45~$\pm$~0.27 &0.47~$\pm$~0.09&43.19~$\pm$~0.21&42.52~$\pm$~0.39&2.08~$\pm$~0.22&2.09~$\pm$~0.21\\
96&J142753.79+345248.35&0.514&1.58~$\pm$~0.11 &0.27~$\pm$~0.05&41.76~$\pm$~0.21&42.12~$\pm$~0.11&0.41~$\pm$~0.27&0.54~$\pm$~0.21\\
97&J142918.15+592106.65&0.739&2.70~$\pm$~0.14 &0.73~$\pm$~0.03&43.59~$\pm$~0.08&43.87~$\pm$~0.12&2.38~$\pm$~0.10&2.48~$\pm$~0.08\\
98&J142943.07+474726.23&0.221&1.27~$\pm$~0.30 &0.94~$\pm$~0.03&43.28~$\pm$~0.08&43.81~$\pm$~0.05&1.96~$\pm$~0.10&2.16~$\pm$~0.08\\
99&J143624.81-002905.36&0.325&1.66~$\pm$~0.31 &0.95~$\pm$~0.05&42.70~$\pm$~0.20&43.30~$\pm$~0.09&1.23~$\pm$~0.29&1.47~$\pm$~0.20\\
100&J144231.82+014353.43&0.280&0.99~$\pm$~0.14 &0.84~$\pm$~0.08&42.36~$\pm$~0.23&42.15~$\pm$~0.23&1.22~$\pm$~0.24&1.25~$\pm$~0.23\\
102&J145108.76+270926.92&0.064&0.63~$\pm$~0.21 &0.31~$\pm$~0.03&41.61~$\pm$~0.07&42.21~$\pm$~0.05&0.26~$\pm$~0.09&0.51~$\pm$~0.07\\
103&J145538.73+002238.06&0.434&1.71~$\pm$~0.21 &0.90~$\pm$~0.01&42.51~$\pm$~0.18&42.86~$\pm$~0.07&1.27~$\pm$~0.25&1.40~$\pm$~0.18\\
104&J152114.26+222743.87&0.136&1.02~$\pm$~0.16 &0.88~$\pm$~0.04&42.51~$\pm$~0.12&42.83~$\pm$~0.06&1.28~$\pm$~0.13&1.39~$\pm$~0.12\\
105&J154530.24+484608.98&0.399&2.51~$\pm$~0.08 &0.48~$\pm$~0.02&42.56~$\pm$~0.09&42.55~$\pm$~0.06&1.42~$\pm$~0.10&1.47~$\pm$~0.09\\
107&J161413.20+260416.21&0.131&0.91~$\pm$~0.26 &0.65~$\pm$~0.02&43.45~$\pm$~0.05&43.90~$\pm$~0.06&2.08~$\pm$~0.06&2.24~$\pm$~0.05\\
108&J163352.34+402115.66&0.782&2.14~$\pm$~0.08 &0.50~$\pm$~0.11&42.72~$\pm$~0.26&42.81~$\pm$~0.33&1.55~$\pm$~0.29&1.61~$\pm$~0.26\\
109&J163915.81+412833.70&0.690&1.84~$\pm$~0.05 &0.45~$\pm$~0.10&42.51~$\pm$~0.32&42.89~$\pm$~0.18&1.26~$\pm$~0.38&1.40~$\pm$~0.32\\
110&J171033.22+584456.86&0.281&1.53~$\pm$~0.15 &0.80~$\pm$~0.07&42.49~$\pm$~0.18&42.36~$\pm$~0.14&1.36~$\pm$~0.19&1.39~$\pm$~0.18\\
111&J171352.43+584201.25&0.521&2.11~$\pm$~0.20 &0.29~$\pm$~0.08&42.58~$\pm$~0.17&43.15~$\pm$~0.13&1.26~$\pm$~0.21&1.48~$\pm$~0.17\\
112&J220759.39+001722.62&0.368&1.40~$\pm$~0.21 &0.21~$\pm$~0.03&41.22~$\pm$~0.13&40.99~$\pm$~0.24&0.03~$\pm$~0.15&0.06~$\pm$~0.13\\
113&J223607.68+134355.32&0.326&1.64~$\pm$~0.13 &0.41~$\pm$~0.13&42.26~$\pm$~0.30&43.15~$\pm$~0.20&0.27~$\pm$~0.61&1.07~$\pm$~0.30\\
114&J233741.32+001743.77&0.762&2.08~$\pm$~0.05 &0.43~$\pm$~0.02&42.47~$\pm$~0.10&42.47~$\pm$~0.11&1.25~$\pm$~0.12&1.30~$\pm$~0.10\\
115&J235156.13-010913.34&0.174&1.46~$\pm$~0.15 &0.26~$\pm$~0.02&42.20~$\pm$~0.05&42.59~$\pm$~0.04&0.96~$\pm$~0.06&1.09~$\pm$~0.05\\
117&PG0003+199&0.025&$-$0.09~$\pm$~0.24 &0.53~$\pm$~0.01&41.81~$\pm$~0.05&42.45~$\pm$~0.02&0.36~$\pm$~0.08&0.64~$\pm$~0.05\\
118&PG0007+106&0.089&0.94~$\pm$~0.17 &0.18~$\pm$~0.03&42.16~$\pm$~0.07&42.82~$\pm$~0.04&0.75~$\pm$~0.09&1.05~$\pm$~0.07\\
119&PG0026+129&0.142&0.49~$\pm$~0.15 &0.24~$\pm$~0.08&42.03~$\pm$~0.17&41.69~$\pm$~0.15&0.89~$\pm$~0.18&0.92~$\pm$~0.17\\
120&PG0043+039&0.384&1.24~$\pm$~0.04 &0.33~$\pm$~0.13&42.44~$\pm$~0.32&42.33~$\pm$~0.32&1.31~$\pm$~0.34&1.35~$\pm$~0.32\\
121&PG0049+171&0.064&$-$0.37~$\pm$~0.29 &0.63~$\pm$~0.02&42.25~$\pm$~0.08&42.99~$\pm$~0.03&0.56~$\pm$~0.14&0.95~$\pm$~0.08\\
123&PG0052+251&0.155&1.13~$\pm$~0.17 &0.53~$\pm$~0.05&42.96~$\pm$~0.11&43.61~$\pm$~0.08&1.57~$\pm$~0.14&1.85~$\pm$~0.11\\
124&PG0157+001&0.164&2.53~$\pm$~0.10 &0.90~$\pm$~0.04&43.09~$\pm$~0.09&43.51~$\pm$~0.06&1.85~$\pm$~0.10&1.99~$\pm$~0.09\\
125&PG0804+761&0.100&0.73~$\pm$~0.27 &0.45~$\pm$~0.04&42.52~$\pm$~0.12&42.57~$\pm$~0.09&1.37~$\pm$~0.13&1.42~$\pm$~0.12\\
128&PG0921+525&0.035&$-$0.32~$\pm$~0.15 &0.79~$\pm$~0.03&42.67~$\pm$~0.06&42.89~$\pm$~0.06&1.42~$\pm$~0.06&1.51~$\pm$~0.06\\
129&PG0923+201&0.190&0.84~$\pm$~0.16 &0.61~$\pm$~0.13&42.32~$\pm$~0.35&42.17~$\pm$~0.22&1.18~$\pm$~0.36&1.21~$\pm$~0.35\\
130&PG0923+129&0.029&0.56~$\pm$~0.16 &0.70~$\pm$~0.04&41.90~$\pm$~0.12&41.94~$\pm$~0.11&0.75~$\pm$~0.14&0.81~$\pm$~0.12\\
131&PG0934+013&0.050&0.47~$\pm$~0.16 &0.84~$\pm$~0.04&42.12~$\pm$~0.08&42.60~$\pm$~0.06&0.76~$\pm$~0.10&0.93~$\pm$~0.08\\
132&PG0947+396&0.206&1.23~$\pm$~0.24 &1.06~$\pm$~0.09&43.30~$\pm$~0.23&43.56~$\pm$~0.14&2.09~$\pm$~0.25&2.19~$\pm$~0.23\\
134&PG1001+054&0.161&0.84~$\pm$~0.26 &0.52~$\pm$~0.03&42.08~$\pm$~0.10&42.54~$\pm$~0.05&0.78~$\pm$~0.13&0.94~$\pm$~0.10\\
135&PG1004+130&0.240&1.68~$\pm$~0.18 &0.13~$\pm$~0.07&41.77~$\pm$~0.25&42.61~$\pm$~0.11&0.06~$\pm$~0.68&0.67~$\pm$~0.25\\
137&PG1012+008&0.185&1.36~$\pm$~0.17 &0.62~$\pm$~0.16&42.95~$\pm$~0.32&43.03~$\pm$~0.24&1.77~$\pm$~0.33&1.84~$\pm$~0.32\\
138&PG1022+519&0.045&0.33~$\pm$~0.04 &0.72~$\pm$~0.11&41.70~$\pm$~0.24&41.89~$\pm$~0.16&0.51~$\pm$~0.25&0.59~$\pm$~0.24\\
139&PG1048+342&0.167&0.90~$\pm$~0.17 &1.45~$\pm$~0.05&43.80~$\pm$~0.13&43.97~$\pm$~0.07&2.62~$\pm$~0.14&2.69~$\pm$~0.13\\
141&PG1049-005&0.357&2.31~$\pm$~0.15 &0.45~$\pm$~0.14&43.28~$\pm$~0.27&43.99~$\pm$~0.20&1.84~$\pm$~0.30&2.19~$\pm$~0.27\\
142&PG1100+772&0.313&1.65~$\pm$~0.20 &0.43~$\pm$~0.13&43.32~$\pm$~0.26&43.90~$\pm$~0.20&1.99~$\pm$~0.29&2.22~$\pm$~0.26\\
143&PG1103-006&0.425&1.36~$\pm$~0.25 &0.54~$\pm$~0.13&43.09~$\pm$~0.36&43.40~$\pm$~0.34&1.81~$\pm$~0.45&1.92~$\pm$~0.36\\
144&PG1114+445&0.144&0.79~$\pm$~0.14 &0.61~$\pm$~0.09&42.55~$\pm$~0.21&43.10~$\pm$~0.16&1.23~$\pm$~0.26&1.44~$\pm$~0.21\\
145&PG1115+407&0.154&1.42~$\pm$~0.04 &0.58~$\pm$~0.03&42.24~$\pm$~0.13&42.59~$\pm$~0.06&0.97~$\pm$~0.15&1.10~$\pm$~0.13\\
146&PG1116+215&0.177&1.16~$\pm$~0.02 &0.45~$\pm$~0.14&42.30~$\pm$~0.49&41.54~$\pm$~1.01&1.15~$\pm$~0.51&1.16~$\pm$~0.49\\
148&PG1121+422&0.234&0.72~$\pm$~0.26 &0.13~$\pm$~0.08&41.60~$\pm$~0.21&42.07~$\pm$~0.26&0.33~$\pm$~0.31&0.50~$\pm$~0.21\\
149&PG1126-041&0.060&0.96~$\pm$~0.22 &0.42~$\pm$~0.13&41.61~$\pm$~0.30&41.78~$\pm$~0.24&0.44~$\pm$~0.33&0.51~$\pm$~0.30\\
150&PG1149-110&0.049&0.43~$\pm$~0.30 &0.22~$\pm$~0.10&41.19~$\pm$~0.20&39.95~$\pm$~5.65&0.07~$\pm$~0.24&0.08~$\pm$~0.20\\
151&PG1151+117&0.176&0.47~$\pm$~0.03 &0.54~$\pm$~0.13&42.26~$\pm$~0.31&42.66~$\pm$~0.26&1.03~$\pm$~0.36&1.17~$\pm$~0.31\\
152&PG1202+281&0.165&1.33~$\pm$~0.16 &0.28~$\pm$~0.03&42.21~$\pm$~0.06&42.71~$\pm$~0.05&0.89~$\pm$~0.07&1.07~$\pm$~0.06\\
154&PG1216+069&0.334&0.95~$\pm$~0.28 &0.21~$\pm$~0.11&42.61~$\pm$~0.35&43.25~$\pm$~0.28&1.22~$\pm$~0.60&1.51~$\pm$~0.35\\
155&PG1226+023&0.158&1.74~$\pm$~0.02 &0.05~$\pm$~0.00&42.50~$\pm$~0.09&41.45~$\pm$~0.45&1.38~$\pm$~0.10&1.39~$\pm$~0.09\\
156&PG1229+204&0.064&0.67~$\pm$~0.25 &0.28~$\pm$~0.03&41.33~$\pm$~0.07&41.57~$\pm$~0.06&0.13~$\pm$~0.08&0.22~$\pm$~0.07\\
160&PG1307+085&0.155&0.88~$\pm$~0.29 &0.16~$\pm$~0.06&42.18~$\pm$~0.14&42.97~$\pm$~0.09&0.57~$\pm$~0.24&1.07~$\pm$~0.14\\
161&PG1309+355&0.184&1.22~$\pm$~0.24 &0.69~$\pm$~0.07&42.72~$\pm$~0.20&43.33~$\pm$~0.12&1.36~$\pm$~0.27&1.61~$\pm$~0.20\\
163&PG1322+659&0.168&1.15~$\pm$~0.10 &0.62~$\pm$~0.13&42.60~$\pm$~0.32&42.96~$\pm$~0.19&1.37~$\pm$~0.34&1.50~$\pm$~0.32\\
164&PG1341+258&0.087&0.48~$\pm$~0.16 &1.07~$\pm$~0.10&42.72~$\pm$~0.23&42.89~$\pm$~0.19&1.55~$\pm$~0.24&1.62~$\pm$~0.23\\
166&PG1351+640&0.087&1.51~$\pm$~0.25 &0.26~$\pm$~0.03&42.17~$\pm$~0.09&42.95~$\pm$~0.05&0.48~$\pm$~0.19&0.94~$\pm$~0.09\\
167&PG1352+183&0.158&0.37~$\pm$~0.15 &0.59~$\pm$~0.13&41.94~$\pm$~0.38&41.75~$\pm$~0.27&0.82~$\pm$~0.40&0.85~$\pm$~0.38\\
169&PG1402+261&0.164&1.55~$\pm$~0.09 &0.63~$\pm$~0.06&42.82~$\pm$~0.18&42.93~$\pm$~0.10&1.61~$\pm$~0.19&1.67~$\pm$~0.18\\
171&PG1411+442&0.089&0.57~$\pm$~0.14 &0.48~$\pm$~0.13&42.08~$\pm$~0.50&42.95~$\pm$~0.20&0.19~$\pm$~1.62&0.92~$\pm$~0.50\\
172&PG1415+451&0.114&0.92~$\pm$~0.33 &0.52~$\pm$~0.09&41.37~$\pm$~0.25&41.97~$\pm$~0.19&0.02~$\pm$~0.36&0.26~$\pm$~0.25\\
173&PG1416-129&0.129&0.70~$\pm$~0.13 &0.23~$\pm$~0.07&42.04~$\pm$~0.14&42.26~$\pm$~0.15&0.80~$\pm$~0.16&0.89~$\pm$~0.14\\
174&PG1425+267&0.366&1.81~$\pm$~0.09 &0.34~$\pm$~0.05&42.80~$\pm$~0.12&43.34~$\pm$~0.09&1.49~$\pm$~0.14&1.70~$\pm$~0.12\\
175&PG1426+015&0.086&1.17~$\pm$~0.17 &0.77~$\pm$~0.09&42.58~$\pm$~0.34&42.44~$\pm$~0.34&1.43~$\pm$~0.38&1.47~$\pm$~0.34\\
176&PG1427+480&0.221&1.13~$\pm$~0.34 &1.06~$\pm$~0.04&43.44~$\pm$~0.10&43.92~$\pm$~0.06&2.13~$\pm$~0.12&2.30~$\pm$~0.10\\

\bottomrule
\end{tabular}
\end{adjustwidth}
\end{table}

\begin{table}[H]\ContinuedFloat
\footnotesize
\caption{{\em Cont.}}
\label{tab:o2sfrresult}
\setlength{\tabcolsep}{2.16mm}

\begin{adjustwidth}{-\extralength}{0cm}

\begin{tabular}{m{1.5cm}<{\raggedleft} m{2.6cm}<{\raggedright} c c c c c c c}

\toprule
\multirow{2}{*}{\textbf{Sample {ID} 
}} & \multirow{2}{*}{\textbf{Name}} & \multirow{2}{*}{\boldmath$z$} &
\boldmath$\log {\rm SFR}_{\rm FIR}$ &
\multirow{2}{*}{\boldmath$E(B-V)_{\rm gas}$} &
\boldmath$\log L_{\rm [O\,II]}^{\rm corr}$ &
\boldmath$\log L_{\rm [O\,III]}^{\rm corr}$ &
\boldmath$\log {\rm SFR}_{\rm [O\,II]}$ &
\boldmath$\log {\rm SFR}_{\rm [O\,II]}^{\rm noAGNcorr}$ \\
&  &  &
\textbf{($M_\odot~{\rm yr}^{-1}$)} &
&
\textbf{(\boldmath${\rm erg~s^{-1}}$) }&
\textbf{(\boldmath${\rm erg~s^{-1}}$)} &
\textbf{(\boldmath$M_\odot~{\rm yr}^{-1}$)} &
\textbf{(\boldmath$M_\odot~{\rm yr}^{-1}$)} \\
\midrule
178&PG1440+356&0.077&1.33~$\pm$~0.08 &0.80~$\pm$~0.10&42.57~$\pm$~0.28&42.37~$\pm$~0.41&1.43~$\pm$~0.31&1.46~$\pm$~0.28\\
180&PG1448+273&0.065&0.51~$\pm$~0.28 &0.23~$\pm$~0.10&41.77~$\pm$~0.24&41.97~$\pm$~0.34&0.60~$\pm$~0.29&0.68~$\pm$~0.24\\
182&PG1512+370&0.371&1.47~$\pm$~0.27 &0.38~$\pm$~0.04&42.87~$\pm$~0.12&43.43~$\pm$~0.08&1.54~$\pm$~0.15&1.76~$\pm$~0.12\\
183&PG1519+226&0.137&1.01~$\pm$~0.22 &0.88~$\pm$~0.08&42.30~$\pm$~0.27&42.62~$\pm$~0.13&1.09~$\pm$~0.32&1.20~$\pm$~0.27\\
186&PG1543+489&0.400&2.35~$\pm$~0.18 &0.47~$\pm$~0.01&42.57~$\pm$~0.08&42.55~$\pm$~0.05&1.42~$\pm$~0.09&1.47~$\pm$~0.08\\
187&PG1545+210&0.266&0.98~$\pm$~0.03 &0.28~$\pm$~0.06&42.74~$\pm$~0.14&43.58~$\pm$~0.09&1.01~$\pm$~0.32&1.62~$\pm$~0.14\\
189&PG1612+261&0.131&0.95~$\pm$~0.31 &0.65~$\pm$~0.02&43.45~$\pm$~0.05&43.90~$\pm$~0.06&2.16~$\pm$~0.07&2.32~$\pm$~0.05\\
190&PG1613+658&0.129&1.76~$\pm$~0.09 &0.38~$\pm$~0.14&42.42~$\pm$~0.29&42.76~$\pm$~0.21&1.19~$\pm$~0.31&1.31~$\pm$~0.29\\
194&PG1704+608&0.371&2.17~$\pm$~0.10 &0.75~$\pm$~0.08&43.93~$\pm$~0.16&44.41~$\pm$~0.12&2.65~$\pm$~0.18&2.82~$\pm$~0.16\\
195&PG2112+059&0.466&2.13~$\pm$~0.28 &0.64~$\pm$~0.11&43.43~$\pm$~0.26&43.52~$\pm$~0.18&2.27~$\pm$~0.27&2.33~$\pm$~0.26\\
196&PG2130+099&0.061&1.01~$\pm$~0.11 &0.65~$\pm$~0.01&42.22~$\pm$~0.05&42.81~$\pm$~0.03&0.78~$\pm$~0.08&1.01~$\pm$~0.05\\
197&PG2209+184&0.070&0.33~$\pm$~0.08 &0.31~$\pm$~0.03&41.56~$\pm$~0.14&41.01~$\pm$~0.67&0.44~$\pm$~0.16&0.45~$\pm$~0.14\\
198&PG2214+139&0.067&0.20~$\pm$~0.11 &0.22~$\pm$~0.04&41.56~$\pm$~0.17&41.59~$\pm$~0.08&0.40~$\pm$~0.19&0.45~$\pm$~0.17\\
199&PG2233+134&0.325&1.55~$\pm$~0.21 &0.43~$\pm$~0.11&42.30~$\pm$~0.27&43.17~$\pm$~0.16&0.38~$\pm$~0.59&1.08~$\pm$~0.27\\
200&PG2251+113&0.323&1.28~$\pm$~0.10 &0.14~$\pm$~0.05&42.69~$\pm$~0.12&43.52~$\pm$~0.08&1.02~$\pm$~0.22&1.58~$\pm$~0.12\\
201&PG2304+042&0.042&$-$0.45~$\pm$~0.18 &0.57~$\pm$~0.09&41.71~$\pm$~0.22&42.38~$\pm$~0.13&0.28~$\pm$~0.28&0.60~$\pm$~0.22\\
202&PG2308+098&0.432&1.29~$\pm$~0.06 &0.38~$\pm$~0.13&42.72~$\pm$~0.33&43.62~$\pm$~0.21&0.74~$\pm$~1.19&1.61~$\pm$~0.33\\
\bottomrule
\end{tabular}

\end{adjustwidth}
\footnotesize
\noindent{Notes:} 
Column (1): sample ID, which follows the original numbering of the full sample of 202 quasars. The IDs are not necessarily consecutive because objects without reliable [O~{\sc ii}] or [O {\sc iii}] measurements are omitted. Column (2): source name. Column (3): redshift. Column (4): FIR SFR derived from the cold-dust component of the SED integrated over 8--1000~\textmu m. Column (5): gas color excess used for the internal extinction correction. Columns (6) and (7): extinction-corrected luminosities of [O~{\sc ii}] $\lambda3727$ and [O {\sc iii}] $\lambda5007$, respectively. Column (8): [O~{\sc ii}] SFR after subtracting the AGN narrow-line-region contribution estimated as $0.109L_{\rm [O\,III]}$. Column (9): [O~{\sc ii}] SFR computed without subtracting this [O {\sc iii}] narrow-line-region contribution.
\end{table}

\vspace{-7pt}






\appendixtitles{yes} 
\appendixstart
\appendix

\begin{adjustwidth}{-\extralength}{0cm}
\printendnotes[custom] 

\reftitle{References}

\PublishersNote{}
\end{adjustwidth}
\end{document}